\documentclass[12pt]{article}
\usepackage{amsfonts,url,fancyhdr,amsmath,amscd,epsfig,psfrag,amsthm}

\def\CY{Calabi--Yau}

\def\CY{Calabi--Yau}
\def\ipo{\hbox{\bf 0}}

\usepackage[hmargin=1in,vmargin={1.3in,1in}]{geometry}

\catcode`\"=\active \let"=\"

\def\ifundefined#1{\expandafter\ifx\csname#1\endcsname\relax}
\def\bye{\end{document}}   
\long\def\new#1\endnew{{\bf #1}}
\long\def\del#1\enddel{} 

\def\HS#1 {\hspace*{#1pt}} \def\VS#1 {\vspace*{#1pt}}
\def\BC{\begin{center}}    
\def\EC{\end{center}}

\let\0=\over      \let\1=\vec      \def\2{{1\over2}}    \let\3=\ss
\let\4=\underline \let\5=\overline \let\6=\partial      \def\7#1{{#1}\llap{/}}
\def\8#1{{\textstyle{#1}}}         \def\9#1{{\ifmmode{\pmb{#1}}\else\bf#1\fi}}

          \def\({\left(}       \def\){\right)}
\def\[{\left[}       \def\]{\right]}

\def\mao#1{\mathop{\rm #1}\nolimits}  
       \def\Tr{\mao{Tr}}       
          
\def\Re{\mao{\hbox{\cal Re}}}   \def\Im{\mao{\hbox{\cal Im}}}

\let\and=\wedge                   
      
\let\hc=\dagger                
\let\bra=\langle        \let\ket=\rangle        \def\<#1\>{\bra #1 \ket}

\let\to=\rightarrow

\def\rel#1 #2{\buildrel #1 \over {#2}}

\let\a=\alpha   \let\b=\beta    \let\g=\gamma   \let\d=\delta   
\let\z=\zeta    \let\h=\eta       
   \let\l=\lambda  \let\m=\mu      
\let\n=\nu            \let\p=\pi      \let\r=\rho     \let\s=\sigma 
\let\t=\tau     \let\o=\omega   \let\c=\chi         
            \let\O=\Omega   \let\S=\Sigma 
        \let\L=\Lambda  \let\G=\Gamma   \let\D=\Delta

   \def\ch{{\cal H}}
  \def\ck{{\cal K}} 
\def\cm{{\cal M}}  \def\co{{\cal O}}

 \let\CM=\cm \let\CO=\co    

\def\IR{{\mathbb R}} \def\IC{{\mathbb C}} \def\IP{{\mathbb P}}
   
\def\IZ{{\mathbb Z}}

\def\BC{\begin{center}}
\def\EC{\end{center}}
\def\eeql#1 {\label{#1}\eeq}      \let\nn=\nonumber  
\def\beq{\begin{equation}}      \def\eeq{\end{equation}}        
\def\bea{\begin{eqnarray}}      \def\eea{\end{eqnarray}}
\def\eeal#1 {\label{#1}\eea} 

\def\refeq#1{(\ref{#1})}

\newtheorem{theorem}{Theorem}[section]

\newtheorem{corollary}[theorem]{Corollary}
\newtheorem{proposition}[theorem]{Proposition}

\theoremstyle{definition}
\newtheorem{definition}[theorem]{Definition}
\newtheorem{example}[theorem]{Example}

\newtheorem*{acknowledgments}{Acknowledgments}

\theoremstyle{remark}
\newtheorem{remark}[theorem]{Remark}

\numberwithin{equation}{section}

\newcommand{\barefootnote}[1] {%
  \begingroup
    \renewcommand{\thefootnote}{$\clubsuit$}
    \footnotemark
    \footnotetext{\ #1}
    \renewcommand{\thefootnote}{\arabic{footnote}}
  \endgroup
}

\renewcommand{\title}[1] {%
  \begingroup
    \begin{center}
      \vspace{2in}
      \bf\Large
      \addtolength{\baselineskip}{5mm}
      #1
    \end{center}
  \endgroup
}

\renewcommand{\author}[1] {%
  \begingroup
    \begin{center}
      \vspace{1in}
      \bf
      #1
      \vspace{0.4in}
    \end{center}
  \endgroup
}

\newcommand{\address}[1] {%
  \begingroup
    \begin{center}
      {\it #1}
      \vspace{1in}
    \end{center}
  \endgroup
}

\renewcommand{\date}[1] {%
  \begingroup
      \vspace{0.5in}
      #1
  \endgroup
}

\newcounter{secondpage}
\begin{document}

\baselineskip=15pt plus 1pt\parskip=7pt

\bibliographystyle{./utcaps}

\title{Lectures on complex geometry, \CY\ manifolds and toric geometry}
\author{Vincent Bouchard\barefootnote{{\url{vbouchard@perimeterinstitute.ca}}}}
\address{Perimeter Institute\\
31 Caroline Street North\\
Waterloo, Ontario\\
Canada N2L 2Y5}

\begin{abstract}
These are introductory lecture notes on complex geometry, \CY\ manifolds and toric geometry. We first define basic concepts of complex and K\"ahler geometry. We then proceed with an analysis of various definitions of \CY\ manifolds. The last section provides a short introduction to toric geometry, aimed at constructing \CY\ manifolds in two different ways; as hypersurfaces in toric varieties and as local toric \CY\ threefolds. These lecture notes supplement a mini-course that was given by the author at the Modave Summer School in Mathematical Physics 2005, and at CERN in 2007. 
\end{abstract}

\date{February 2007}

\newpage

\baselineskip=15pt plus 1pt\parskip=1pt

\tableofcontents

\newpage

\baselineskip=15pt plus 1pt\parskip=7pt

\noindent{\bf General remarks}

In these lectures I assume a basic knowledge of differential geometry and vector bundles on real manifolds. If needed the reader may want to consult for instance \cite{Nakahara:2002}.

The two first sections use what I would call a `bundle' approach to complex geometry, following closely the treatment of \cite{Joyce:2000} and the first chapters of \cite{Hori:2003}. For a more traditional approach (from a physicist's point of view), I would recommend the lectures on complex geometry by Philip Candelas \cite{Candelas:1987is} and Nakahara's book \cite{Nakahara:2002}. 

In the third section I define \CY\ manifolds and consider in details two examples of \CY\ threefolds. A good reference (for mathematically-oriented physicists) on \CY\ manifolds is \cite{Hubsch:1992}.

Section 4 is almost independent of the three other sections. It provides a quick introduction to toric geometry, focusing on constructing \CY\ manifolds in toric geometry. It is clearly not self-contained and should be seen as a complement to standard references in toric geometry such as \cite{Skarke:1998yk,Greene:1996cy,Fulton:1993}. This section is mainly based on the second chapter of \cite{Bouchard:2005th}.

I do not pretend to add anything new to these standard mathematical topics in these lecture notes; in fact, various definitions and propositions have been borrowed almost literally from the references cited above. Rather, the aim is to provide an opportunity for students and researchers in mathematical physics to get a grip on these mathematical concepts without having to go through the standard lengthy books. I also tried to build a bridge between the mathematical and physical expositions on these subjects. However, for the sake of brevity most proofs are omitted, and various interesting topics are not even discussed. 

Indeed, it is obvious that a proper introduction to the subject of these lecture notes requires much more than a four hour crash course. Therefore, I had to make some choices in the topics covered; my selection was mainly dictated by applications in string theory.

Finally, if the reader is not sure about the meaning of certain concepts, I recommend doing a search on \url{http://wikipedia.org/}. And don't forget that Wikipedia is a collaborative project; add your own comments and definitions if needed!

\begin{acknowledgments} I would like to thank Philip Candelas, Wen Jiang, Marcos Mari\~no and Fonger Ypma for interesting discussions while writing these lecture notes. I owe special thanks to Wen Jiang for proof reading these notes before the Modave school. My work was partly supported by a Rhodes Scholarship and an NSERC PGS Doctoral (and Postdoctoral) Fellowship.
\end{acknowledgments}

\newpage

\section{Complex geometry}

In this section we define complex manifolds, develop calculus on them and define a few important topological invariants. We assume that the reader is familiar with real manifolds; hence, this section may be understood as providing the complex analogs to the familiar concepts of real geometry.

\subsection{Complex manifolds}\label{complex}

Let us first define complex manifolds in two different ways. For a third definition involving principal bundles, see \cite{Joyce:2000}.

\begin{definition}\label{d:compl1}
Let $M$ be a real $2m$-dimensional manifold and $\{U_i\}$ an open covering on $M$. On each open subset $U_i$, we define a {\it coordinate chart} to be the pair $(U_i,\psi_i)$ where $\psi_i:U \to \IC^m$ is an homeomorphism from $U_i$ to an open subset of $\IC^m$ (that is $\psi_i$ gives a set of complex coordinates $z^i_1,\ldots,z^i_m$ on $U_i$). We say that the data $(M,\{U_i,\psi_i\})$ is a {\it complex manifold} if for every non-empty intersections $U_i \cap U_j$ the transition functions $\psi_{ij}=\psi_j \circ \psi_i^{-1}: \psi_i (U_i \cap U_j) \to \psi_j (U_i \cap U_j)$ are holomorphic as maps from $\IC^m$ to itself (i.e. they depend only on the $z_{\m}$ but not on their complex conjugate). $m$ is called the {\it complex dimension} of a complex manifold.
\end{definition}

\begin{figure}[htp]
\begin{center}
\psfrag{U_i}{$U_i$}
\psfrag{U_j}{$U_j$}
\psfrag{IC^m}{$\IC^m$}
\psfrag{p_i}{$\psi_i$}
\psfrag{p_j}{$\psi_j$}
\psfrag{M}{$M$}
\includegraphics[width=8cm]{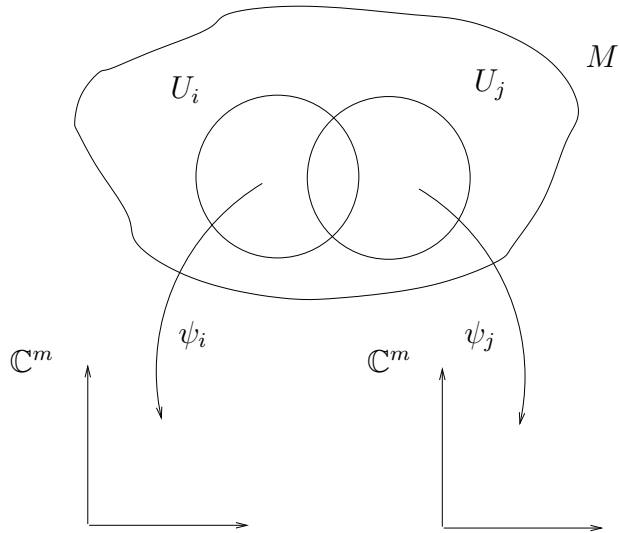}
\caption{Pictorial representation of a complex manifold, according to definition \ref{d:compl1}.}
\label{complexfig}
\end{center}
\end{figure}

Thus, roughly speaking, a complex manifold is a topological space that locally looks like $\IC^m$. By the definition above we see that a complex manifold is always a real manifold. However, the converse is not always true: when does a real manifold can be seen as a complex manifold, that is when does a real manifold admit a complex structure? The second definition of complex manifolds answers this interesting question.

But let us first fix some notation.

\begin{remark}
In the following, we will always use latin indices for real coordinates and greek indices for complex coordinates. Moreover, $z_{\bar{\mu}}$ will be a shorthand for $\bar{z}_{\bar{\mu}}$, where $\bar{\mu}$ is a normal tensor index.
\end{remark}

\begin{remark}
To fix notation: let $M$ be a real $n$-manifold, with tangent bundle $TM$ and cotangent bundle $T^*M$. A smooth section $S$ of the tensor product bundle $\otimes^k TM \otimes^l T^*M$ is called a {\it tensor field of type $(k,l)$} on $M$, and its components in a local coordinate basis are denoted by $S_{a_1,\ldots,a_l}^{b_1,\ldots,b_k}$. The space of tensor fields (the space of sections) of type $(k,l)$ on $M$ is denoted by $\G (\otimes^k TM \otimes^l T^*M)$. In fact, given any vector bundle $E$, we will always denote its space of sections by $\G(E)$.
\end{remark}

Let $M$ be a real $2m$-dimensional manifold. We define an {\it almost complex structure} $J$ on $M$ to be a smooth tensor field $J \in \G(TM \otimes T^*M)$ on $M$ satisfying $J_a^b J_b^c = -\delta_a^c$. 

Now take a vector field $v \in \G (TM)$, with components $v^a$ in a coordinate basis, and use the fact that $J$ is a map on tangent spaces to define a new vector field $(Jv)^b = J_a^b v^a$. Since $J_a^b J_b^c = -\delta_a^c$, we see that $J(Jv) = -v$, that is $J^2 = -1$. Therefore, roughly speaking $J$ is a generalization of the usual multiplication by $\pm i$ in complex analysis; it gives to each tangent space $T_p M$ to a point $p$ in $M$ the structure of a complex vector space. We say that a real $2m$-dimensional manifold endowed with an almost complex structure $J$ is an {\it almost complex manifold}.

For two vector fields $v,w$, define now a vector field $N_J(v,w)$ by
\beq
N_J(v,w) = [v,w]+J[v,Jw]+J[Jv,w]-[Jv,Jw],
\eeq
where $[,]$ denotes the Lie bracket of vector fields.\footnote{The Lie bracket $[v,w]$ of two vector fields given in a coordinate basis by $v^a {\partial_a}$ and $w^b {\partial_b}$ is defined by $(v^a \partial_a w^b-w^a \partial_a v^b) \partial_b.$} $N$ is called the {\it Nijenhuis tensor}. In local coordinates it is given by
\beq
N^a_{bc} = J^d_b (\partial_d J^a_c - \partial_c J^a_d) - J^d_c (\partial_d J^a_b-\partial_b J^a_d).
\eeq

We can now state our second definition of complex manifolds.

\begin{definition}
Let $M$ be a $2m$-dimensional real manifold and $J$ an almost complex structure on $M$. If $N \equiv 0$, we call $J$ a {\it complex structure} on $M$. A {\it complex manifold} is defined by the data $(M,J)$ where $J$ is a complex structure on $M$.
\end{definition}

It follows from a theorem by Newlander and Nirenberg (which we will not prove here) that the two definitions are equivalent. Namely, the theorem states that it is possible to find local complex coordinates with holomorphic transition functions if and only if the almost complex structure is integrable, which is equivalent to say that its Nijenhuis tensor vanishes. 

Therefore, a real manifold can be considered as a complex manifold only if it admits a complex structure $J$. Let us now pause to give a few examples of complex manifolds.

\begin{example}
The simplest example of a complex manifold of complex dimension $m$ is $\IC^m$, which obviously admits a global coordinate chart.
\end{example}

\begin{example}
A very important family of complex manifolds are the {\it complex projective spaces}, denoted by $\IC \IP^m$ (often written simply as $\IP^m$). These are the spaces of complex lines through the origin in $\IC^{m+1}$. Take the space $\IC^{m+1} \backslash \{0\}$ and quotient by the identification
\beq\label{e:proj}
(z_0,\ldots,z_m) \sim \lambda (z_0,\ldots,z_m),
\eeq
where $\lambda$ is any non-zero complex number. We call the $z_{\mu}$ {\rm homogeneous coordinates} on $\IC \IP^m$. One way to show that it is a complex manifold is to define a set of coordinate charts with holomorphic transition functions. Since the $z_{\mu}$ are not all zero, we can choose an open covering defined by the open subsets
\beq
U_{\a} = \{ z_{\a} \neq 0 \}. 
\eeq
On each $U_{\a}$ we define coordinates $\z_{\mu}^{\a} = z_{\mu} / z_{\a}$, which we call {\rm inhomogeneous coordinates} (or {\rm affine coordinates}). On the intersection $U_{\alpha} \cap U_{\beta}$ we have that
\beq
\z_{\m}^{\a} = {z_{\m} \over z_{\a}} = {z_{\m} \over z_{\b}} {z_{\b} \over z_{\a}} = {\z_{\m}^{\b} \over \z_{\b}^{\a}},
\eeq
since both $z_{\a}$ and $z_{\b}$ are non-zero on the intersection. Therefore, the transition function $\psi_{\a \b}^{(\m)}: \z^{\a}_{\m} (U_{\a} \cap U_{\b}) \to \z^{\b}_{\m} (U_{\a} \cap U_{\b})$ is simply a multiplication by $(\z^{\a}_{\b})^{-1}$, which is of course holomorphic. The manifolds $\IC \IP^m$ are also compact, which we will not prove here.
\end{example}

\begin{example}
Let us now consider submanifolds of the above complex manifolds. If, for the moment, we restrict our attention to compact complex manifolds, we see that submanifolds of $\IC^m$ are not very interesting, since a theorem that we will not prove here (see \cite{Candelas:1987is} for a simple proof) states that a connected compact analytic submanifold of $\IC^m$ is a point. However, many compact complex manifolds can be constructed as submanifolds of projective spaces $\IC \IP^m$. We saw that $\IC \IP^m$ is compact; all its closed complex submanifolds are also compact. In fact, there is a theorem by Chow that states that all such submanifolds of $\IC \IP^m$ can be realized as the zero locus of a finite number of homogeneous polynomial equations (which means polynomial equations homogeneous in the homogeneous coordinates $z_{\m}$). One important example is the Fermat quintic in $\IC \IP^4$, given as the zero locus of the equation
\beq
\sum_{\m=0}^4 (z_{\m})^5 = 0.
\eeq
This three-dimensional compact complex manifold turns out to be \CY, probably the most studied \CY\ threefold. We will come back to this manifold in section \ref{quintic}. 

One can also generalize the construction above by considering the zero locus of a finite number of homogeneous polynomial equations in a product of projective spaces, rather than a single projective space. This generalization leads to a large number of manifolds, many of them \CY. Complex manifolds constructed as the zero locus of a set of polynomial equations are usually called {\rm complete intersections}. We will give some examples of these in section \ref{CICY}.
\end{example}

\begin{example}
Let us consider, as a final example in this section, a simple generalization of complex projective spaces: {\it weighted projective spaces}. Basically, instead of quotienting by a $\IC^*$ action acting as in \eqref{e:proj}, one assigns a different weight to each coordinate of $\IC^{m+1}$. For instance, the two-dimensional weighted projective space $\IC \IP^{(2,3,1)}$ can be constructed by taking the space $\IC^{3} \setminus \{ 0 \}$ and quotienting by the $\IC^*$ action
\beq
(z_0,z_1,z_2) \sim  (\l^2 z_0, \l^3 z_1, \l z_2),
\eeq
where $\l$ is a non-zero complex number; the superscripts $(2,3,1)$ denote the weights of the $\IC^*$ action. It is easy to show, as for projective spaces, that this is a compact complex manifold. Moreover, one can consider as in the previous example hypersurfaces and complete intersections in weighted projective spaces and products thereof.
\end{example}

\subsection{Tensors on complex manifolds}

In the previous section we introduced the notion of a complex structure $J$, which can be understood as an endomorphism of real tangent spaces: at a point $p$ of $M$, $J$ gives a linear map $J: T_pM \to T_pM$. Let us now complexify the tangent space $T_pM$ to get $T_pM \otimes \IC$, which is a complex vector space isomorphic to $\IC^{2m}$. The map $J_p$ extends naturally to a map $J:T_pM \otimes \IC \to T_pM \otimes \IC$.

Since $J^2=-{1}$, the eigenvalues of $J$ in $T_pM \otimes \IC$ are $\pm i$. Now let $T^{(1,0)}_p M$ (resp. $T^{(0,1)}_p M$) denote the eigenspace of $J_p$ with eigenvalue $i$ (resp. $-i$). Both eigenspaces are isomorphic to $\IC^m$, complex conjugate to each other, and we have the decomposition $T_pM \otimes \IC = T^{(1,0)}_p M \oplus T^{(0,1)}_p M$. Since this works at any point $p$ of $M$, it can be extended to the full bundle $TM$; that is, the complexified tangent bundle (which we will now denote by $T_{\IC} M$) decomposes as $T_{\IC} M = T^{(1,0)} M \oplus T^{(0,1)} M$. We call $T^{(1,0)} M$ (resp. $T^{(0,1)} M$) the {\it holomorphic (resp. anti-holomorphic) tangent bundle}. In some way, by complexifying the tangent bundle and using the action of the complex structure, we trade the real tangent bundle $TM$ for the holomorphic tangent bundle $T^{(1,0)} M$.

\begin{remark}
As noted above to obtain this decomposition we must complexify the tangent bundle, that is its sections are given by complex-valued vector fields on a complex manifold. It is important to note the difference between having a complex manifold and complexifying the vector bundles, which is not the same thing. In principle, we could consider real vector bundles on complex manifolds, which would lead to real-valued tensor fields on complex manifolds, or complex vector bundles on real manifolds, which would give complex-valued tensor fields on real manifolds.
\end{remark}

The decomposition into holomorphic and anti-holomorphic pieces carries through to complexified cotangent bundles as well, hence $T_{\IC}^* M = T^{*(1,0)} M \oplus T^{*(0,1)} M$. Now, since a tensor field is defined as a section of tensor products of tangent and cotangent bundles, we expect to find a similar decomposition for complex-valued tensor fields on complex manifolds.

\begin{remark}
We remarked earlier that $z_{\bar{\mu}}$ was a shorthand for $\bar{z}_{\bar \mu}$. Similarly, for general tensors, the indices $\bar{\a},\bar{\b},\bar{\m},\ldots$ are tensor indices, but they indicate a modification to the tensor itself.
\end{remark}

Let us define the tensors $S^{\a \ldots}_{\ldots} = {1 \over 2}(S^{a \ldots}_{\ldots}-i J^a_m S^{m \ldots}_{\ldots})$ and $S^{\bar{\a} \ldots}_{\ldots} = {1 \over 2}(S^{a \ldots}_{\ldots}+i J^a_m S^{m \ldots}_{\ldots})$, and similarly  $T^{\ldots}_{\b \ldots} = {1 \over 2}(T^{\ldots}_{b \ldots}-i J^m_b T^{\ldots}_{m \ldots})$ and $T^{ \ldots}_{\bar{\b} \ldots} = {1 \over 2}(T^{\ldots}_{b \ldots}+i J_b^m T^{\ldots}_{m \ldots})$. It is clear that $S^{a \ldots}_{\ldots}=S^{\a \ldots}_{\ldots}+S^{\bar{\a} \ldots}_{\ldots}$ and $T^{\ldots}_{b \ldots}=T^{\ldots}_{\b \ldots}+T^{ \ldots}_{\bar{\b} \ldots}$. Therefore, these operations on tensors are projections: the `$\a$' operation projects on the holomorphic piece, while the `$\bar{\a}$' operation projects on the anti-holomorphic piece. 

Moreover, it is easy to show that in this notation $J^a_b = i\delta_{\b}^{\a} -i \delta_{\bar{\b}}^{\bar{\a}}$, which means that $J$ acts on tensor indices $\a,\b,\ldots$ (resp. $\bar{\a},\bar{\b},\ldots$) by multiplication by $i$ (resp. $-i$). Thus, tensors (at a point $p$) with indices $\a,\b,\ldots$ lie in a tensor product of the holomorphic tangent and cotangent spaces, while tensors with indices $\bar{\a},\bar{\b},\ldots$ lie in a tensor product of the anti-holomorphic tangent and cotangent spaces. In particular, a vector $v^{\a}$ is called a {\it holomorphic vector}, while a vector $v^{\bar{\a}}$ is called an {\it anti-holomorphic vector}, and we have the decomposition $v^a = v^{\a}+v^{\bar{\a}}$. This yields the desired decomposition of tensor fields into holomorphic and anti-holomorphic parts.

\subsection{Exterior forms on complex manifolds}

Let us first recall some properties of exterior forms on real manifolds.

\begin{definition}
Let $M$ be a real $n$-dimensional manifold. The $k$-th exterior (or wedge) power of the cotangent bundle $T^*M$ (defined by the exterior product of its sections below) is written $\L^k T^*M$. Smooth sections of $\L^k T^*M$ are called {\it $k$-forms}, and the vector space of $k$-forms is denoted by $\G(\L^k T^*M)$ or $\O^k (M)$.
\end{definition}

Thus, $r$-forms are totally antisymmetric tensor fields of type $(0,r)$. We can define the {\it exterior product} (or wedge product) $\wedge$ and the {\it exterior derivative} $d$ of $k$-forms as follows. Let $\a$ be a $k$-form and $\b$ be a $l$-form, then $\a \wedge \b$ is a $(k+l)$-form and $d\a$ is a $(k+1)$-form defined by (in component notation)
\bea
(\a \wedge \b)_{a_1 \ldots a_{k+l}} &=& \a_{[a_1 \ldots a_k} \b_{a_{k+1} \ldots a_{k+l}]},\\
(d \a)_{a_1 \ldots a_k} &=& \partial_{[a_1} \a_{a_2 \ldots a_{k+1}]},
\eea
where $[\ldots]$ denotes complete antisymmetrization.

It follows that
\beq
d(d\a)=0,~~~~\a \wedge \b = (-1)^{kl} \b \wedge \a,~~~~d(\a \wedge \b) = (d\a) \wedge \b + (-1)^k \a \wedge (d\b).
\eeq

The first property is usually denoted by $d^2=0$. We say that a $r$-form $\a$ is {\it closed} if it satisfies $d \a =0$, and {\it exact} if it can be written as $\a = d \b$ for some $(r-1)$-form. Since $d^2=0$, any exact form is closed. 

Now we want to study exterior forms on complex manifolds. Using the decomposition of the complexified cotangent bundle, it is easy to show that the following complexified bundles decompose as
\beq\label{formdecomp}
\L^k T^*_{\IC} M = \bigoplus_{j=0}^k \L^{j,k-j} M,
\eeq
where we defined $\L^{p,q} M := \L^p T^{*(1,0)} M \otimes \L^{q} T^{*(0,1)} M$. A section of $\L^{p,q} M$ is called a {\it $(p,q)$-form}, which is a complex-valued differential form with $p$ holomorphic pieces and $q$ anti-holomorphic pieces. We denote the vector space of $(p,q)$-forms by $\G (\L^{p,q} M)$ or $\O^{p,q}(M)$, and the vector space of complexified $k$-forms by $\G(\L^k T^*_{\IC} M)$ or $\O^k_\IC (M)$.

On a complex manifold, the exterior derivative also admits a simple decomposition: $d = \partial + \bar{\partial}$, where we defined the operators $\partial: \O^{p,q}(M) \to \O^{p+1,q}(M)$ and $\bar{\partial}: \O^{p,q}(M) \to \O^{p,q+1}(M)$. The identity $d^2=0$ implies that $\partial^2=\bar{\partial}^2=0$ and $\partial \bar{\partial}+\bar{\partial} \partial=0$. We can also define a real operator $d^c: \O^k_{\IC} (M) \to \O^{k+1}_{\IC} (M)$ by $d^c = i (\bar{\6}-\6)$, which satisfies
\beq
d d^c + d^c d=0,~~~(d^c)^2=0,~~~\6 = \2 (d+id^c),~~~\bar{\6} = \2 (d-id^c),~~~d d^c = 2i \6 \bar{\6}.
\eeq

\subsection{Cohomology}

Cohomology is a very important part of geometry. To motivate the introduction of cohomology, let us make an analogy with gauge theories in physics. 

In a physical gauge theory, gauge transformations correspond to mathematical transformations that do not change the physics. Hence, any physical observable must be gauge invariant. Another way of seeing this is to define equivalence classes, by which we mean classes of objects that only differ by gauge transformations. Observables could then be defined as equivalence classes, since by definition these classes are gauge invariant.

In geometry, we are interested in finding fundamental properties of geometrical systems. Hence, we also want to define some objects that are invariant under `uninteresting' transformations, analogous to gauge transformations in physics. This is precisely what cohomology theories provide.

For instance, topological invariants are of prime importance in the study of manifolds. There are various topological invariants that one can define on a manifold; cohomology theories and characteristic classes (for instance Chern classes) provide many of them. In fact, even from a physics point of view topological invariants are important: many physical questions can be reformulated as questions about topological invariants of manifolds, for example in compactifications of string theory and in supersymmetry.

Let us start by giving a broad definition of the concept of cohomology groups.

\begin{definition}
Let $A_0,A_1,\ldots$ be abelian groups connected by homomorphisms $d_n:A_n \to A_{n+1}$, such that the composition of two consecutive maps is zero: $d_{n+1} \circ d_n = 0$ for all $n$. We can then form the {\it cochain complex}
\beq
\begin{array}{ccccccccc}
&d_0&&d_1&&d_2&&d_3\\
0 &\to& A_0 &\to& A_1 &\to& A_2 &\to& \ldots
\end{array}
\eeq
The {\it cohomology groups} $H^k$ are defined by
\beq
H^k = {{\rm Ker}(d_k: A_k \to A_{k+1}) \over {\rm Im} (d_{k-1}: A_{k-1} \to A_k)}.
\eeq 
\end{definition}

The elements of $H^k$ are the equivalence classes alluded to above, consisting of elements in ${\rm Ker(d_k)}$ that only differ by a transformation in ${\rm Im (d_{k-1})}$.

With this general definition of cohomology groups, defining a particular cohomology boils down to finding a collection of abelian groups and homomorphisms such that the above definition holds. The first cohomology theory that we will look at is defined over real manifolds.

\begin{definition}
Let $M$ be a real $n$-dimensional manifold. As $d^2=0$, we can form the complex
\beq
\begin{array}{ccccccccccc}
&d&&d&&d&&d&&d\\
0 &\to& \O^0 (M)&\to& \O^1 (M)&\to& \ldots &\to& \O^n (M) &\to&0.
\end{array}
\eeq
We define the {\it de Rham cohomology groups} $H^k_{dR} (M, \IR)$ of $M$ by
\beq
H^k_{dR} (M,\IR) = {{\rm Ker}(d: \O^k(M) \to \O^{k+1}(M)) \over {\rm Im} (d: \O^{k-1}(M) \to \O^k(M))}.
\eeq
\end{definition}

In other words, $H^k_{dR} (M, \IR)$ is the set of closed $k$-forms where two forms are considered equivalent if they differ by an exact form, i.e. $\omega \simeq \omega + d\alpha$; it is the quotient of the vector space of closed $k$-forms on $M$ by the vector space of exact $k$-forms on $M$. That is, given a closed $k$-form $\omega$, its {\it cohomology class} $[\omega] \in H^k_{dR} (M, \IR)$ is the space of closed $k$-forms which differ from $\omega$ by an exact form. We call $\omega$ a {\it representative} of $[\omega]$.

We can see in the definition of de Rham cohomology groups that the complex terminates. This is because there is no antisymmetric $(n+1)$-tensor field on an $n$-manifold.

\begin{remark}
The $\IR$ in $H^k_{dR} (M,\IR)$ means that the closed $k$-forms are real, i.e. they are elements of $\O^k (M)$. However, we can also define the de Rham cohomology groups for complexified $k$-forms, that is elements of $\O^k_\IC (M)$, which we denote by $H^k_{dR} (M, \IC)$.
\end{remark}

Using cohomology groups we can define some important topological invariants. We define the {\it Betti numbers} $b^k = \dim_\IR H^k_{dR}(M, \IR)$. We can also define the {\it Euler characteristic} as the alternating sum of the Betti numbers:
\beq
\c = \sum_{k=0}^n (-1)^k b^k.
\eeq

Now what is the analog of the de Rham cohomology groups for complex manifolds? 

\begin{definition}
Let $M$ be a complex manifold of complex dimension $m$. As $\bar{\6}^2=0$, we can form the complex
\beq
\begin{array}{ccccccccccc}
&\bar{\6}&&\bar{\6}&&\bar{\6}&&\bar{\6}&&\bar{\6}&\\
0 &\to& \O^{p,0}(M) &\to& \O^{p,1}(M) &\to& \ldots &\to& \O^{p,m}(M) &\to&0.
\end{array}
\eeq
We define the {\it Dolbeault cohomology groups} $H^{p,q}_{\bar{\6}} (M)$ of $M$ by
\beq
H^{p,q}_{\bar{\6}} (M) = {{\rm Ker}(\bar{\6}: \O^{p,q}(M) \to \O^{p,q+1}(M)) \over {\rm Im} (\bar{\6}: \O^{p,q-1}(M) \to \O^{p,q}(M))}.
\eeq
\end{definition}

Remark that the Dolbeault cohomology groups depend on the complex structure of $M$. Note also that we could have defined the cohomology groups using $\6$ instead of $\bar{\6}$, this is just a matter of convention since they are complex conjugate. 

We now define the {\it Hodge numbers}  to be $h^{p,q} = \dim H^{p,q}_{\bar{\6}} (M)$. The Hodge numbers of a complex manifold are summarized in what is commonly called the {\it Hodge diamond}:
\beq
\begin{array}{ccccc}
&&h^{m,m}&& \\
&h^{m,m-1}&\vdots&h^{m-1,m}& \\
h^{m,0}&\cdots&&\cdots&h^{0,m}\\
&h^{1,0}&\vdots&h^{1,0}& \\
&&h^{0,0}&& \\
\end{array}
\eeq

The $(m+1)^2$ Hodge numbers are not independent; there are many relations between them, depending on the kind of complex manifold we are looking at. We will investigate some of them in sections 2 and 3.

There are many other cohomology groups that one can define on a manifold, for instance the well-known \v{C}ech cohomology groups; they are all isomorphic on smooth manifolds. For the purpose of these lectures the de Rham and Dolbeault cohomology groups will suffice.

\subsection{Chern classes}\label{chern}

Given a fiber $F$, a structure group\footnote{The structure group $G$ of a fiber bundle $E$ is a Lie group acting on the left on the fibers $F$. Roughly speaking, the transition functions on overlapping coordinate charts take values in the structure group $G$.} $G$ and a base space $M$, we may construct many fiber bundles over $M$, depending on the choice of transition functions. It is an interesting problem to classify these bundles and see how much they differ from the trivial bundle $M \times F$. In order to do so, characteristic classes are what we are looking for; they are subsets of cohomology classes of the base space which measure the non-triviality, or twisting, of the bundle. In other words, they are `obstructions' which prevent a bundle from being a trivial bundle.

In this section we focus on a particular kind of characteristic classes, namely Chern classes, using their differential geometry definition (for a purely topological definition see for instance \cite{Hori:2003}).

Much more than what we present here could be said about characteristic classes. For instance, we could have discussed in more details the Chern character and Todd classes, which have important applications in physics; in particular for computing the index of some physically relevant operators using the Atiyah-Singer index theorem. But four hours is not enough to enter into this subject; more information on these topics can be found for instance in \cite{Nakahara:2002,Hori:2003,Green:1987mn,Griffiths:1978}.

For the sake of brevity we will not give here the general definition of characteristic classes in terms of invariants polynomials and cohomology classes; the reader is referred to chapter 11 of \cite{Nakahara:2002}. It may be an interesting exercise to show that Chern classes are indeed characteristic classes according to the general definition.

\begin{definition}
Let $E$ be a complex vector bundle over a manifold $M$, and let $F=dA + A \wedge A$ be the curvature two-form of a connection $A$ on $E$. We define the {\it total Chern class} $c(E)$ of $E$ by
\beq
c(E) = \det (1+{i \0 2 \p} F).
\eeq
Since $F$ is a two-form, $c(E)$ is a direct sum of forms of even degrees. We define the {\it Chern classes} $c_k(E) \in H^{2k} (M, \IR)$ by the expansion of $c(E)$:
\beq
c(E)=1+c_1(E)+c_2(E)+\ldots.
\eeq
\end{definition}

\begin{remark}
To be more precise, the $c_k(E)$ in the expansion are called the {\it Chern forms}, which are closed $(2k)$-forms, while the Chern classes are defined as the cohomology classes of the Chern forms. Therefore, Chern classes are cohomology classes, while the Chern forms are representatives of the Chern classes.
\end{remark}

This definition relies on a connection $A$ on the bundle, so one may think that Chern classes depend on a choice of connection $A$. Fortunately this is not the case. For different curvatures $F$ and $F'$, the difference between the two invariant polynomials is an exact form, that is the two invariant polynomials are in the same cohomology class. Since the cohomology classes defined by invariant polynomials form what is called characteristic classes, in the present case Chern classes, it follows that Chern classes are independent of the choice of connection, but that different connections will lead to different representatives of the cohomology classes $c_k$. A complete proof of this fact is given in chapter 11 of \cite{Nakahara:2002}.

Since $F$ is a two-form, on an $n$-dimensional manifold the Chern classes $c_j (E)$ with $2j > n$ vanish identically. Also, irrespective of the dimension of $M$, the series terminates at $c_k(E) = \det (iF / 2\p)$ and $c_j(E)=0$ for $j>k$. Therefore $c_j(E)=0$ for $j>k$ where $k$ is the rank of the bundle $E$.

\begin{remark}
When the complex bundle $E$ is the holomorphic tangent bundle $T^{(1,0)} M$, we say that $c_k(E)$ is the Chern class of the manifold $M$ and usually denote it simply by $c_k(M)$ or $c_k$.
\end{remark}

It is useful for computations to have explicit formulae for the Chern classes. One can show that (see \cite{Nakahara:2002}):
\bea\label{chernexplicit}
c_0(E) &=& \[ 1\], \nonumber\\
c_1(E) &=& \[ {i \0 2 \p} \Tr F \],\nonumber\\
c_2(E) &=& \[ \2 \({i \0 2 \p}\)^2 (\Tr F \wedge \Tr F - \Tr (F \wedge F)) \],\nonumber\\
&\vdots&\nonumber\\
c_k (E) &=& \[ \({i \0 2 \p}\)^k \det F \],
\eea
where $\[ \ldots \]$ denotes the cohomology class.

We now give an alternative description of Chern classes that is useful in understanding their topological meaning, using cycles.\footnote{This description comes from the duality between homology and cohomology. However, we will not discuss homology in these lectures; see for instance \cite{Hori:2003,Candelas:1987is} for more information on that topic.} Let $E$ be a rank $r$ complex vector bundle on a manifold $M$ of complex dimension $m$. Let $s_1,\ldots,s_r$ be $r$ global sections of $E$ (not necessarily holomorphic). Define $D_k$ to be the locus of points where the first $k$ sections develop a linear dependence, that is where $s_1 \wedge \ldots \wedge s_k = 0$ as a section of $\L^k E$. Then the cycles $D_k$ are Poincar\'e dual to the Chern classes $c_{r+1-k}$ (see section \ref{quintic} for a definition of Poincar\'e duality, and \cite{Hori:2003} for a more general discussion). Thus we can use the cycles $D_k$ to understand the topological meaning of the Chern classes $c_{r+1-k}$. For instance, the Chern class $c_1 (E)$ corresponds to the cycle $D_r$, which is defined by $s_1 \wedge \ldots \wedge s_r = 0$. This represents the zeroes of the sections of the determinant line bundle $\L^r E$ (see section \ref{holbundle}); therefore $c_1 (E) =0$ is the same thing as $\L^r E$ being trivial. Indeed, $c_1 (E)= c_1 (\L^r E)$. For $k=1$, we find that the top Chern class $c_r(E)$ is represented by the zeros of a single section of $E$. In particular, if $E = T^{(1,0)} M$, then $c_m(M)$ represents the zeros of a generic section of the holomorphic tangent bundle; and the integral of the top Chern class over $M$ gives the Euler characteristic of $M$:
\beq
\c = \int_M c_m (M).
\eeq

\subsubsection{Properties of Chern classes --- and a digression on exact sequences}

Chern classes also have a few important properties that are useful in practical calculations. To state some of them, let us first introduce the notion of (short) exact sequences, which is omnipresent in algebraic geometry.

A sequence of spaces and maps 
\beq
\ldots \to X_1 \xrightarrow{\a_1} X_2 \xrightarrow{\a_2} X_3 \xrightarrow{\a_3} \ldots
\eeq 
such that ${\rm Im} (\a_k) = {\rm Ker} (\a_{k+1})$ for all $k$ is called {\it an exact sequence}. In particular, $0 \to A \xrightarrow{\a} B$ means that $\a$ is injective, while $B \xrightarrow{\b} C \to 0$ means that $\b$ is surjective. 

If the exact sequence has the form
\beq\label{e:ses}
0 \to A \xrightarrow{\a} B \xrightarrow{\b} C \to 0,
\eeq
we call it {\it a short exact sequence}. From the definition above it is easy to see that a short exact sequence is equivalent to saying that $A \subseteq B$ and that $C =  B / A $. We say that the sequence {\it splits} if $B = A \oplus C$. Hence, one can understand the space $B$ in the short exact sequence \eqref{e:ses} as a deformation of the direct sum $A \oplus C$.

\begin{figure}[htp]
\begin{center}
\psfrag{An exact sequence}{An exact sequence}
\psfrag{A short exact sequence}{A short exact sequence}
\psfrag{A}{$A$}
\psfrag{B}{$B$}
\psfrag{C}{$C$}
\psfrag{D}{$D$}
\psfrag{E}{$E$}
\psfrag{0}{$0$}
\includegraphics[width=8cm]{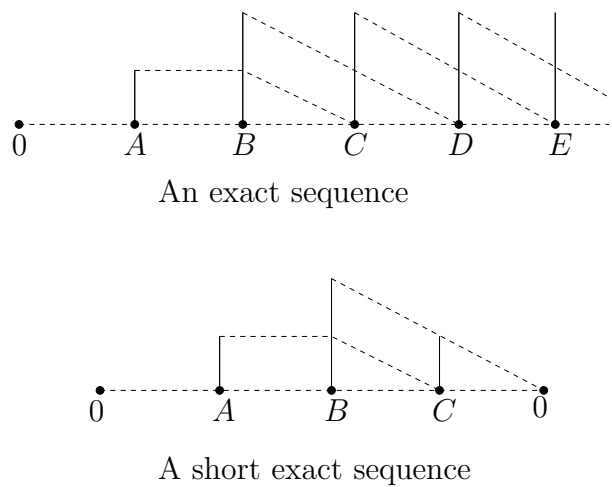}
\caption{Schematic visualization of exact sequences. The vertical lines represent the vectors spaces, the dots represent the zeroes, and the dashed lines represent the maps. This pictorial representation helped me visualize the condition that ${\rm Im} (\a_k) = {\rm Ker} (\a_{k+1})$ for all $k$. And from this picture it is easy to see that giving a short exact sequence, as in the second picture, is equivalent to saying that $A \subseteq B$ and $C = B/A$.}
\label{f:es}
\end{center}
\end{figure}

\begin{remark}
When I first learned about exact sequences I found it easier to understand these definitions pictorially; hence I reproduce in figure \ref{f:es} my schematic visualizations of these abstract definitions. They may or may not be useful to the reader, but well here they are. :-)
\end{remark}

Now let $V$ be the direct sum bundle $E \oplus F$. Then the total Chern class of $V$ is $c(V) = c(E) \wedge c(F)$, which follows from properties of the determinant. In fact, this property is (in an appropriate sense) `deformation independent', and $c(V) = c(E) \wedge c(F)$ holds whenever $0 \to E \to V \to F \to 0$ is a short exact sequence.

\subsubsection{Chern character}

To end this section we quickly define the Chern character, which will be useful in computations of Chern classes later on. Suppose that we define the classes $x_i$ intrinsically by $c(E)=\prod_{i=1}^{r} (1+x_i)$, where $r$ is the rank of the bundle $E$. Then the Chern character $ch(E)$ is defined by $ch(E) = \sum_i e^{x_i}$. The first few terms in the expansion of the exponential are
\beq
ch(E)=r+c_1(E) + \2 (c_1(E)^2 - 2 c_2 (E) ) + {1\over 6} (c_1(E)^3-3c_1(E) c_2(E)+3c_3(E)) + \ldots.
\eeq
Note that the Chern character satisfies the useful identities $ch(E \oplus F) = ch(E)+ch(F)$ and $ch(E \otimes F) = ch(E) ch(F)$.

\subsection{Holomorphic vector bundles}\label{holbundle}

So far we have met complex vector bundles over complex manifolds. We will now formally define holomorphic vector bundles over complex manifolds.

\begin{figure}[htp]
\begin{center}
\psfrag{f}{$f$}
\psfrag{phi}{$\phi$}
\psfrag{IC^m}{$\IC^m$}
\psfrag{IC^n}{$\IC^n$}
\psfrag{psi}{$\psi$}
\psfrag{M}{$M$}
\psfrag{N}{$N$}
\psfrag{U}{$U$}
\psfrag{V}{$V$}
\includegraphics[width=10cm]{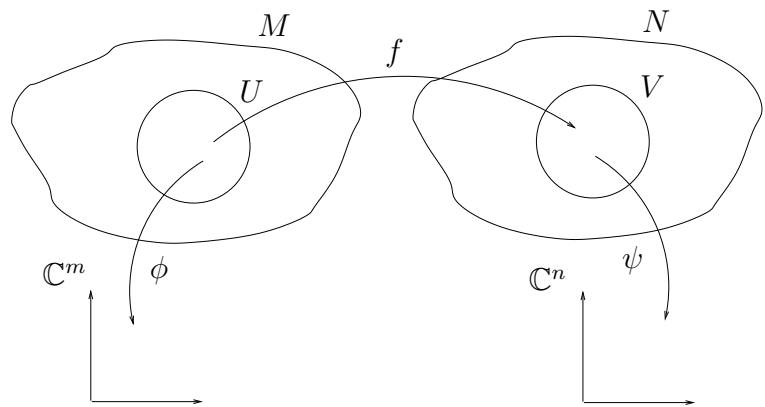}
\caption{Pictorial representation of a holomorphic map of complex manifolds.}
\label{holomormapfig}
\end{center}
\end{figure}

\begin{definition}
Let $M$ be a complex manifold. Let $\{E_p\}: p \in M$ be a family of complex vector spaces of dimension $k$, parameterized by $M$. Let $E$ be the total space of this family, and $\p : E \to M$ be the natural projection. Suppose also that $E$ has the structure of a complex manifold. $E$ with its complex structure is called a {\it holomorphic vector bundle} with fiber $\IC^k$ if the map $\p$ is a holomorphic map of complex manifolds\footnote{Let $f:M\to N$, and $M$ and $N$ be complex manifolds with complex dimensions $m$ and $n$. Take a point $p$ in a chart $(U,\phi)$ of $M$. Let $(V,\psi)$ be a chart of $N$ such that $f(p) \in V$. Let $\{z_{\mu}\} = \phi(p)$ and $\{ w_{\nu}\} = \psi(f(p))$ be their coordinates in $\IC^m$ and $\IC^n$. We thus have a map $\psi \circ f \circ \phi^{-1}: \IC^m \to \IC^n$. If each function $w_{\nu}$ is a holomorphic function of $z_{\mu}$, we say that $f$ is a {\it holomorphic map}. A map $f$ is called {\it biholomorphic} if an inverse map $f^{-1}: N \to M$ exists and both $f$ and $f^{-1}$ are holomorphic maps. See figure \ref{holomormapfig} for a pictorial description of holomorphic maps of complex manifolds.} and for each $p$ there exists an open neighborhood $U \subset M$ and a biholomorphic map $\phi_U: \p^{-1} (U) \to U \times \IC^k$ such that for each $u \in U$ the map $\phi_U$ takes $E_u$ to $\{u\} \times \IC^k$, and this is an isomorphism between $E_u$ and $\IC^k$ as complex vector spaces. $k$ is called the {\it rank} of the bundle; it is the dimension of its fibers.
\end{definition}

\begin{figure}[htp]
\begin{center}
\psfrag{times}{$\times~U $}
\psfrag{phi_U}{$\phi_U$}
\psfrag{IC^m}{$\IC^k$}
\psfrag{pi}{$\pi$}
\psfrag{E}{$E$}
\psfrag{M}{$M$}
\psfrag{P}{$p$}
\psfrag{U}{$U$}
\includegraphics[width=8cm]{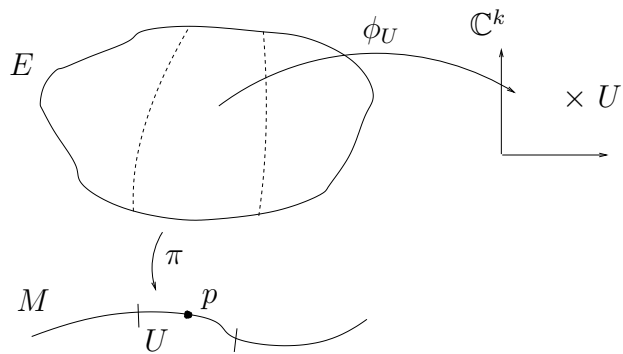}
\caption{Pictorial representation of a holomorphic vector bundle.}
\label{holomorphicfig}
\end{center}
\end{figure}

Basically, the important points in the definition are that the total space $E$ has the structure of a complex manifold, the projection map $\pi$ is a holomorphic map of complex manifolds, and the trivialization $\phi_U$ is a biholormophism. This is not always obvious, and not all complex vector bundles are holomorphic vector bundles.

Let us now give a few examples of holomorphic vector bundles. The simplest holomorphic vector bundle is $M \times \IC^k$, which is called the {\it trivial} vector bundle over $M$. 

The complexified tangent and cotangent bundles also admit natural complex structures, which make them into holomorphic vector bundles.

Now what about the complex vector bundles $\L^{p,q} M$? It turns out that, although these bundles all have complex vector spaces as fibers, only the bundles with $q=0$ are holomorphic vector bundles. A holomorphic section of $\L^{p,0} M$ is called a {\it holomorphic $p$-form}. 

\subsubsection{Holomorphic line bundles}

If the fiber of a holomorphic vector bundle is $\IC$, i.e. its rank is one, then we say that it is a {\it holomorphic line bundle}. As an example, the bundle $K_M = \L^{m,0} M$ on a complex manifold $M$ of complex dimension $m$, sections of which are $(m,0)$-forms, is a holomorphic line bundle, usually called the {\it canonical bundle}. Its sections are sometimes called holomorphic volume forms on $M$. Actually, given any holomorphic vector bundle $E$ of rank $r$, we can form the holomorphic line bundle $\L^r E$, the {\it determinant line bundle}, whose transition functions are the determinants of those for $E$. 

Given two line bundle $L$ and $L'$ over $M$, one can construct many other line bundles. First, there is the dual line bundle $L^*$ to $L$. But also, any tensor product of line bundles is also a line bundle, so $L \otimes L'$ forms a new line bundle. This is so because if we look at the fibers, they are one-dimensional vector spaces; but we have that $\dim (U \otimes V) = \dim U \dim V$ for vector spaces $U$ and $V$, so we see that the fibers of $L \otimes L'$ also have dimension one, therefore $L \otimes L'$ is a line bundle. In fact, the set of isomorphism classes of holomorphic line bundles over $M$ form an abelian group, where multiplication is given by the tensor product, inverses are dual bundles, and the identity is the trivial line bundle $L \otimes L^*$. This group is called the {\it Picard group} of $M$.

Now let us consider the special case where $M$ is $\IC \IP^m$. First, there is the natural line bundle whose fiber over a point $l$ in $\IC \IP^m$ is the line it represents in $\IC^{m+1}$; this is the {\it tautological line bundle} (also called the {\it universal line bundle}), which we denote by $L^{-1}$. Its dual, which we denote by $L$, is called the {\it hyperplane line bundle}. In fact, since the tensor product of two holomorphic line bundles is always a holomorphic line bundle, we can construct holomorphic line bundles $L^k$ over $\IC \IP^m$ for any $k \in \IZ$. Actually, it can even be shown that every holomorphic line bundle over $\IC \IP^m$ is isomorphic to $L^k$ for some $k \in \IZ$. For instance, the canonical bundle $K_{\IC \IP^m}$ is isomorphic to $L^{-m-1}$. By abuse of notation, we usually denote the line bundle $L^k$ by $\co(k)$, although technically $\co(k)$ denotes the sheaf of holomorphic sections of $L^k$. It is interesting to note that for $k \geq 0$ the vector space of holomorphic sections $\G(L^k)$ is canonically identified with the set of homogeneous polynomials of degree $k$ in $\IC \IP^{m}$. Therefore, homogeneous coordinates of $\IC \IP^{m}$ are sections of the hyperplane line bundle $L$. We will come back to this correspondence in the next subsection.

\subsubsection{Refined Dolbeault cohomology groups}

Finally, we can use these results to refine our definition of the Dolbeault cohomology groups on complex manifolds. First, we must generalize our definition of a $k$-form. We defined $k$-forms as sections of the bundle $\L^k M$, that is they take values in $M$. We can now define an $E$-valued $k$-form $\o$, that is a form that takes values in the vector bundle $E$, by the map $\o : E \to E \otimes \L^k M$. Thus $\o$ is a section of $E \otimes \L^k M$. For instance, we could define a $\L^{p,0} M$-valued $(0,q)$-form $\o: \L^{p,0} M \to \L^{p,0} M \otimes \L^{0,q} M$; hence $\o$ is a section of $\L^{p,q} M$, that is $\o$ is simply what we previously defined as a $(p,q)$-form. But using this generalized definition we can consider forms taking values in complex vector bundles $E$ different from $\L^{p,0} M$.

We defined the $(p,q)$ Dolbeault cohomology group as the quotient of the vector space of $\bar{\6}$-closed  $(p,q)$-forms by the vector space of $\bar{\6}$-exact $(p,q)$-forms. However, the operator $\bar{\6}$ only acts on the `anti-holomorphic' piece of a $(p,q)$-form, i.e. on the $(0,1)$ part of the decomposition of the $\L^{p,q} M$ bundle. Therefore, we can look at the elements of the Dolbeault cohomology groups in the following way; they are $\bar{\6}$-closed $(0,q)$-forms taking values in the holomorphic vector bundle $\L^{p,0} M$. From that point of view, we can define the cohomology groups $H_{\bar{\6}}^q ( \L^{p,0} M)$, where $\L^{p,0} M$ is a holomorphic vector bundle, sections of which are $(p,0)$-forms. These cohomology groups are indeed isomorphic to the Dolbeault cohomology groups previously defined. However, they can be generalized; since $\bar{\6}$ commutes with holomorphic transition functions, $\bar{\6}$ may now act on forms taking values in any holomorphic vector bundle $E$, not just the holomorphic vector bundle $\L^{p,0}M$. This leads to the following generalized definition of Dolbeault cohomology.

\begin{definition}
Let $M$ be a complex manifold, and $E$ a holomorphic vector bundle on $M$. We have the complex
\beq
0 \to \G (E) \xrightarrow{\bar{\6}} \G(E) \otimes \O^{0,1} (M) \xrightarrow{\bar{\6}} \ldots\xrightarrow{\bar{\6}} \G(E) \otimes \O^{0,m} (M) \xrightarrow{\bar{\6}} 0.
\eeq
We define the {\it Dolbeault cohomology groups taking values in $E$}, $H^{q}_{\bar{\6}} (E)$, by
\beq
H^{q}_{\bar{\6}} (E) = {{\rm Ker}(\bar{\6}: \G(E) \otimes \O^{0,q}(M) \to \G(E) \otimes \O^{0,q+1}(M)) \over {\rm Im} (\bar{\6}: \G(E) \otimes \O^{0,q-1}(M) \to \G(E) \otimes \O^{0,q}(M))}.
\eeq
\end{definition}

This definition reduces to the former definition of Dolbeault cohomology groups when $E=\L^{p,0} M$.

\subsection{Divisors and line bundles}

To end this section, we explore the connection between line bundles and divisors, which are important objects in algebraic geometry and in its physical applications. In practice, it is convenient to be able to think of divisors in terms of line bundles and vice-versa. For more on that topic see \cite{Griffiths:1978}.

\begin{definition}
Let $M$ be a complex manifold. Let $N$ be an hypersurface, that is a codimension $1$ submanifold that can be written locally as the zero locus of an holomorphic function. Moreover, let $N$ be irreducible, that is it cannot be written as the union of two hypersurfaces. Then we define a {\it divisor} to be the formal finite sum of irreducible hypersurfaces $D=\sum_i n_i N_i$ with integer coefficients $n_i$. A divisor is called {\it effective} if $n_i \geq 0 $ for all $i$.
\end{definition}

Now in which way is this object related to line bundles?

Let $L$ be a holomorphic line bundle over $M$, and $s$ a nonzero holomorphic section of $L$. Let $N$ be the hypersurface defined by $N= \{ m \in M : s(m)=0 \}$. Then $N$ may be written in a unique way as a union $N = \cup_i N_i$, where $N_i$ are irreducible hypersurfaces. For each $i$, there is a unique positive integer $a_i$ which tells us the order to which $s$ vanishes along $N_i$. Define $D=\sum_i a_i N_i$. Then $D$ is an effective divisor. In fact, since the divisors constructed that way from line bundles are equal if and only if the line bundles are isomorphic, there is a bijection between effective divisors on $M$ and isomorphism classes of holomorphic line bundles equipped with nonzero holomorphic sections.

If we now consider holomorphic line bundles with a nonzero meromorphic\footnote{A meromorphic function is a function that is holomorphic on an open subset of the complex number plane $\IC$ except at points in a set of isolated poles, which are certain well-behaved singularities. Every meromorphic function can be expressed as the ratio between two holomorphic functions (with the denominator not constant $0$): the poles then occur at the zeroes of the denominator.} section, then the divisor $D=\sum_i a_i N_i$ corresponds to a section $s$ with a zero of order $a_i$ along $N_i$ if $a_i > 0$, and a pole of order $-a_i$ along $N_i$ if $a_i < 0$. Thus there is a bijection between divisors on $M$ and isomorphism classes of holomorphic line bundles equipped with nonzero meromorphic sections.

\section{K\"ahler geometry}

We will now consider a special type of complex manifolds, namely K\"ahler manifolds. Roughly speaking, manifolds with a K\"ahler metric are those for which the parallel transport of a holomorphic vector remains holomorphic. 

\subsection{K\"ahler manifolds}\label{kahlerdef}

\begin{definition}
Let $(M,J)$ be a complex manifold, and let $g$ be a Riemannian metric on $M$. We call $g$ a {\it Hermitian metric} if the three following equivalent conditions hold:
\begin{enumerate}
\item{$g(v,w) = g(Jv,Jw)$ for all vector fields $v,w$ on $M$;}
\item{In component notation, $g_{ab} = J_a^c J_b^d g_{cd}$;}
\item{Using the greek indices notation, $g_{ab}=g_{\a \bar{\b}}+g_{\bar{\a} \b}$, that is $g_{\a \b} = g_{\bar{\a} \bar{\b}} = 0$.}
\end{enumerate}
\end{definition}

In other words, a Hermitian metric is a positive-definite inner product $T^{(1,0)}M \otimes T^{(0,1)} M \to \IC$ at every point on a complex manifold $M$. We leave it as an exercise for the reader to show that the three above conditions are equivalent.

\begin{remark}
Note that this is a restriction on the metric, not on the manifold $M$, since it can be shown that a complex manifold always admits a Hermitian metric.
\end{remark}

Using this Hermitian metric $g$, we can define a two-form $\o$ on $M$ called the {\it Hermitian form} by $\o(v,w) = g(Jv, w)$ for all vector fields on $v,w$ on $M$. The equivalent definition in terms of real components is $\o_{ab} = J_a^c g_{cb}$, while in terms of complex components it is given by $\o_{ab} = i g_{\a \bar{\b}} - i g_{\bar{\a} \b}$. Therefore, $\o$ is a $(1,1)$-form. 

If $g$ was not Hermitian, then $\o$ would not be a form (that is it would not be antisymmetric). In fact, the Hermitian condition is equivalent to the condition $\o_{ab}=-\o_{ba}$.

\begin{definition}
Let $(M,J)$ be a complex manifold, and $g$ a Hermitian metric on $M$, with Hermitian form $\o$. $g$ is a {\it K\"ahler metric} if $d \o = 0 $. In this case we call $\o$ a {\it K\"ahler form}, and we call a complex manifold $(M,J)$ endowed with a K\"ahler metric a {\it K\"ahler manifold}.
\end{definition}

In the remaining of this section we will explore properties of K\"ahler manifolds. First, it can be shown that locally, the K\"ahler condition $d \o = 0$ is equivalent to the condition $\6_{\m} g_{\n \bar{\a}} = \6_{\n} g_{\m \bar{\a}}$ and its conjugate equation $\6_{\bar{\r}} g_{\m \bar{\a}} = \6_{\bar{\a}} g_{\m \bar{\r}}$. Moreover, for a K\"ahler metric $g$ the Levi-Civita connection has no mixed indices, meaning that vectors with holomorphic indices remain with holomorphic indices after parallel transportation. Hence, for K\"ahler manifolds parallel transport preserves holomorphicity. This implies a restriction on the holonomy of K\"ahler manifolds, as we will see in section \ref{KahlerHolonomy}.

An important consequence of K\"ahlerity is the existence of a K\"ahler potential. Let $\phi$ be a real smooth function on $M$. Clearly, $d d^c \phi$ is a closed real two-form, as both $d$ an $d^c$ are real operators. But since $d d^c = 2i \6 \bar{\6}$, $d d^c \phi$ is also a closed $(1,1)$-form. 

In fact, any closed $(1,1)$-form $\o$ can be expressed {\it locally} as $\o = d d^c \phi$ for a real smooth function $\phi$ on $M$. This is however not true globally; it only holds if $\o$ is also exact.

Therefore, it is always possible to express the K\"ahler form $\o$ in terms of a smooth function $\phi$ locally, and we call this function the {\it K\"ahler potential}. This also follows from the K\"ahler condition in component notation given above. However this is not true globally, for the following reason.

Let $M$ be a compact manifold of real dimension $2m$, with K\"ahler form $\o$. Since $\o$ is closed, it defines a Dolbeault cohomology class $[\o] \in H^{1,1}_{\bar{\6}} (M)$, or a de Rham cohomology class $[\o] \in H^2_{dR} (M, \IR)$. The latter is usually called the {\it K\"ahler class}. Further, the wedge product of $m$ copies of $\o$, denoted by $\o^m$, is proportional to the volume form of $g$ (one can sees that by working out the explicit expression of $\o^m$ using the component definition of $\o_{ab}$). Therefore it defines a non-trivial element in both $H^{m,m}_{\bar{\6}} (M)$ and $H^{2m}_{dR} (M,\IR)$, and we have that $\int_M \o^m  \propto {\rm vol}(M)$. But for a compact manifold $M$, ${\rm vol}(M) > 0$, and $\int_M \o^m$ only depends on the cohomology class $[\o]$ (by Stoke's theorem). Thus, $[\o]$ must be non-zero. However, $d d^c \phi$ is exact and therefore zero as a cohomology class. It follows that on a compact K\"ahler manifold it is impossible to find a globally defined K\"ahler potential.

However, an interesting result is that we can parameterize K\"ahler metrics with a fixed K\"ahler class by smooth functions on the manifold. This goes as follows. Let $\o$ and $\o'$ be two different K\"ahler forms in the same K\"ahler class. Therefore, $\o-\o'$ is an exact form. But we saw that exact forms can be expressed {\it globally}  as $d d^c \phi$ for a smooth function $\phi$. Therefore, globally we have that $\o = \o' + d d^c \phi$. Moreover, $\phi$ is unique up to the addition of a constant; suppose $\phi_1$ and $\phi_2$ are two different such functions, then $d d^c (\phi_1 - \phi_2) = 0$ on $M$, which implies that $\phi_1 - \phi_2$ is constant, as $M$ is compact. Therefore, smooth functions on $M$ parameterize the set of K\"ahler forms in a particular K\"ahler class.

\begin{example}
In this example we sketch the proof that $\IC \IP^m$ is a K\"ahler manifold, that is it admits a K\"ahler metric. Consider the function $u(z_0,\ldots,z_m) = \sum_{\m=0}^m |z_\m|^2$ where $z_\m$, $\m=0,\ldots,m$ are homogeneous coordinates on $\IC^{m+1} \backslash \{0\}$. Define a $(1,1)$-form $\a$ by $\a = d d^c (\log u)$. $\a$ cannot be the K\"ahler form of any metric on $\IC^{m+1} \backslash \{0\}$, since it is not positive. However, if we consider the projection $\pi: \IC^{m+1} \backslash \{0\} \to \IC \IP^m$ defined by $\pi: (z_0,\ldots,z_m) \mapsto [z_0,\ldots,z_m]$, one can show that there exists a unique positive $(1,1)$-form $\o$ on $\IC \IP^m$ such that $\a = \pi^* (\o)$. $\o$ is a K\"ahler form on $\IC \IP^m$; its associated K\"ahler metric is called the {\it Fubini-Study metric}, and is given in components by $g_{\m \bar{\n}} = \6_\m \6_{\bar{\n}} \log u$.
\end{example}

There is a general result that says that any submanifolds of a K\"ahler manifold is also K\"ahler (since the restriction of the K\"ahler form to a complex submanifold is also a closed, positive $(1,1)$-form). We just saw that $\IC \IP^m$ is K\"ahler; therefore all its submanifolds are also K\"ahler. This important family of complex manifolds was constructed in section \ref{complex}; we now know that they all admit a K\"ahler metric.

\subsection{Forms on K\"ahler manifolds}

A complex manifold with a K\"ahler metric has now enough structure to define operators analog to the Hodge star and the $d^{\hc}$ and $\Delta_d$ operators on a Riemannian manifold, thus leading to a Hodge theory on K\"ahler manifolds. We can also relate this theory to the real version --- since complex manifolds are also real manifolds --- and find additional properties of K\"ahler manifolds.

Before doing so let us summarize quickly some results for real manifolds. 

Let $M$ be a compact Riemannian $n$-manifold, with metric $g$. Let $\a$ and $\b$ be $k$-forms on $M$. We define the {\it pointwise inner product} of $\a$ and $\b$ by (in component notation) $(\a,\b) = \a_{a_1 \ldots a_k} \b_{b_1 \ldots b_k} g^{a_1 b_1} \ldots g^{a_k b_k}$. We define a second inner product, called the {\it $L^2$ inner product}, using the volume form $dV_g$ on $M$ given by the metric, by $\<\a,\b\> = \int_M (\a,\b) dV_g$. 

The {\it Hodge star} is an isomorphism of vector bundles $*: \L^k T^*M \to \L^{n-k} T^*M$ such that if $\b$ is a $k$-form on $M$, then $*\b$ is the unique $(n-k)$-form satisfying $\a \wedge (*\b) = (\a, \b) dV_g$ for all $k$-forms $\a$ on $M$.

We define an operator $d^{\hc}: \O^k (M) \to \O^{k-1} (M)$ by $d^{\hc}\b = (-1)^{kn+n+1} * d (*\b)$. It is sometimes called the {\it formal adjoint} of $d$. We have that $(d^{\hc})^2 =0$, so we say that a form satisfying $d^{\hc}\a$ is {\it coclosed}, and if $\a = d^{\hc} \b$ then it is {\it coexact}. We also define the Laplacian $\D_d = d d^{\hc} + d^{\hc} d$. If a form satisfies $\D_d \a = 0$, we say that it is {\it harmonic}. It can be shown easily that a form on a compact manifold $M$ is harmonic if and only if it is closed and coclosed.

Now, let $\ch^k$ be the vector space of harmonic $k$-forms on $M$. The Hodge decomposition theorem states that $\O^k (M) = \ch^k \oplus {\rm Im}(d_{k-1}) \oplus {\rm Im} (d^{\hc}_{k+1})$, that is every $k$-form can be expressed uniquely as a sum of an harmonic form, an exact form and a coexact form. Moreover, we have that ${\rm Ker} (d_k) = \ch^k \oplus {\rm Im}(d_{k-1})$ and ${\rm Ker} (d^{\hc}_k) = \ch^k \oplus {\rm Im}(d^{\hc}_{k+1})$, that is closed (resp. coclosed) forms can be expressed uniquely as a sum of an harmonic and an exact (resp. coexact) form. Therefore, since $H_{dR}^k (M, \IR) = {\rm Ker} (d_k) / {\rm Im}(d_{k-1})$, we see that there is an isomorphism between $\ch^k$ and $H_{dR}^k (M, \IR)$. In other words, every de Rham cohomology class on $M$ contains a unique harmonic representative. However, although cohomology classes are topological invariants of $M$, their harmonic representatives depend on a particular choice of a metric $g$.

We are now ready to see what the analogs of these constructions are on K\"ahler manifolds.

Let $M$ be a complex manifold or real dimension $2m$, with a K\"ahler metric $g$. Let $\a$ and $\b$ be complex $k$-forms on $M$. First, define a pointwise inner product (in real component notation) by $(\a,\b) = \a_{a_1 \ldots a_k} \overline{\b_{b_1 \ldots b_k}} g^{a_1 b_1} \ldots g^{a_k b_k}$, that is $(\a,\b)$ is a complex function on $M$, which is bilinear in $\a$ and $\bar{\b}$. For a compact $M$, define the $L^2$ inner product by $\< \a, \b \> = \int_M (\a,\b) dV_g$. $\< \a, \b \>$ is a complex number, bilinear in $\a$ and $\bar{\b}$.

Let the {\it Hodge star on K\"ahler manifolds} be the isomorphism of complex vector bundles $*: \L^k T_{\IC}^*M \to \L^{2m-k} T_{\IC}^*M$ such that if $\b$ is a complex $k$-form on $M$, then $* \b$ is the unique complex $(2m-k)$-form satisfying $\a \wedge (*\b) = (\a,\b)dV_g$ for all complex $k$-forms $\a$. Define the following operators taking complex $k$-forms to complex $(k-1)$-forms (there is no $(-1)$ factor as in the real analog since the real dimension of a complex manifold is always even):
\beq
d^{\hc} = -*d(*\a),~~~~ \6^{\hc} = -*\6 (*\a),~~~~\bar{\6}^{\hc} = -*\bar{\6}(*\a).
\eeq
We also define the usual Laplacian $\D_d=d d^{\hc}+d^{\hc}d$, and the two Laplacians $\D_{\6} = \6 \6^{\hc} + \6^{\hc} \6$ and $\D_{\bar{\6}} = \bar{\6} \bar{\6}^{\hc} +\bar{\6}^{\hc}\bar{\6}$, which satisfy $\D_{\6} = \D_{\bar{\6}} = \2 \D_d$. Reformulating in terms of holomorphic and anti-holomorphic parts, we have defined the following operators on K\"ahler manifolds:
\bea
*&:&  \O^{p,q} (M) \to  \O^{m-p,m-q} (M);\nonumber\\
\6&:& \O^{p,q}(M) \to \O^{p+1,q}(M),~~~~\bar{\6}: \O^{p,q}(M) \to \O^{p,q+1}(M);\nonumber\\
\6^{\hc}&:& \O^{p,q}(M) \to \O^{p-1,q}(M),~~~~\bar{\6}^{\hc}: \O^{p,q}(M) \to \O^{p,q-1}(M);\nonumber\\
\D_{\6}&:& \O^{p,q}(M) \to \O^{p,q}(M),~~~~\D_{\bar{\6}}: \O^{p,q}(M) \to \O^{p,q}(M).
\eea

\begin{remark}
In the literature the $\bar{\6}$-Laplacian $\D_{\bar{\6}}$ on a K\"ahler manifold is often simply called the Laplacian and denoted by $\D$. Similarly, we call a $(p,q)$-form satisfying $\D_{\bar{\6}}$ an {\it harmonic $(p,q)$-form}. A $(p,q)$-form $\a$ is harmonic if and only if $\6 \a = \bar{\6} \a = \6^{\hc} \a = \bar{\6}^{\hc} \a = 0$.
\end{remark}

How can we define the Hodge theory of a K\"ahler manifold? In fact, we can formulate a Hodge theory for the $\bar{\6}$ operator which is very similar to the Hodge theory for real manifolds.

Let $\ch^{p,q}$ be the vector space of harmonic $(p,q)$-forms. The following decompositions hold:
\bea
\O^{p,q}(M) &=& \ch^{p,q} \oplus \bar{\6} \[ \O^{p,q-1} (M) \] \oplus \bar{\6}^{\hc} \[ \O^{p,q+1} (M) \],\nonumber\\
{\rm Ker} \bar{\6} &=& \ch^{p,q} \oplus \bar{\6} \[ \O^{p,q-1} (M) \],\nonumber\\
 {\rm Ker} \bar{\6}^{\hc} &=& \ch^{p,q} \oplus \bar{\6}^{\hc} \[ \O^{p,q+1} (M) \].
\eea
We see that the vector space of harmonic $(p,q)$-forms is isomorphic to the Dolbeault cohomology groups, that is $\ch^{p,q} \cong H^{p,q}_{\bar{\6}}(M)$. This means that there is a unique harmonic representative in each Dolbeault cohomology classes.

Let us now define $\ch^k_\IC$ to be the vector space of complex harmonic $k$-forms (with respect to $\D_d$), that is $\ch^k_\IC = {\rm Ker} \( \D_d: \O^k_{\IC} (M) \to \O^{k}_{\IC} (M) \)$. Using (\ref{formdecomp}), and the fact that $\2 \D_d = \D_{\bar{\6}}$, we see that there is a further decomposition
\beq
\ch^k_\IC = \bigoplus_{j=0}^k \ch^{j,k-j},
\eeq
of complex harmonic $k$-forms into a sum of harmonic $(p,q)$-forms with $p+q=k$. By the above isomorphisms, we learn that for a K\"ahler manifold, the complexified de Rham cohomology decomposes into the Dolbeault cohomology (note that the de Rham cohomology groups are complex, as we are now considering complexified $k$-forms)
\beq\label{kahlerdecomp}
H^k_{dR} (M, \IC) = \bigoplus_{j=0}^k H^{j,k-j}_{\bar{\6}} (M).
\eeq

This relation and others lead to various properties of cohomology groups on K\"ahler manifolds, which is the topic of the next section.

\subsection{Cohomology}

In this section we explore the relations between cohomology groups on K\"ahler manifolds.

First, using (\ref{kahlerdecomp}) we see directly that for a K\"ahler manifold,
\beq
b^k = \sum_{j=0}^{k} h^{j,k-j}.
\eeq
Note that this is true because $\dim_\IC H^k_{dR} (M, \IC) = \dim_\IR H^k_{dR} (M, \IR)$, since we previously defined the Betti numbers for real de Rham cohomology groups. Moreover, the Hodge star on K\"ahler manifolds and complex conjugation tell us that
\beq
h^{p,q} = h^{q,p},~~~~h^{p,q} = h^{m-q,m-p}.
\eeq
Furthermore, since by the above properties $b^{2k-1} = \sum_{j=0}^{2k-1} h^{j,2k-1-j} = 2 \sum_{j=0}^{k-1} h^{j,2k-1-j}$, we have that $b^{2k-1}$ is even for $1 \leq k \leq m$. In practice, this last condition is useful to deduce that some complex manifolds do not admit a K\"ahler metric. For example, the manifold $S^3 \times S^1$ has a complex structure, but it does not admit a K\"ahler metric since $b^1 =1 $.

It would be interesting to understand what the Hodge numbers of a K\"ahler manifold exactly mean in terms of geometry. To end this subsection, let us make a first step towards this goal by defining the K\"ahler cone.

\begin{definition}
Let $(M,J)$ be a complex manifold admitting K\"ahler metrics. If $g$ is a K\"ahler metric on $M$, then the K\"ahler form $\o$ is a closed $(1,1)$-form, that is $[\o] \in H^{1,1}_{\bar{\6}} (M)$. We define the {\it K\"ahler cone} $\ck$ of $M$ to be the set of cohomology classes $[\o] \in H^{1,1}_{\bar{\6}} (M)$ such that $\o$ is the K\"ahler form of a K\"ahler metric on $M$.
\end{definition}

In other words, the K\"ahler cone defines the set of possible K\"ahler forms on $M$. This points towards the fact that the Hodge number $h^{1,1}(M)$ is intimately related to the K\"ahler structure moduli space of $M$; we will come back to this in the next section.

\subsection{Holonomy}\label{KahlerHolonomy}

Before we close this section, let us find what the holonomy of a K\"ahler manifold is. We have not discussed holonomy of manifolds so far, but they are crucial in the definition of \CY\ manifolds. Let us first recall what the holonomy of a manifold is.

\begin{definition}
Let $M$ be an $n$-dimensional Riemannian manifold with metric $g$ and affine connection $\nabla$. Let $p$ be a point in $M$ and consider the set of closed loops at $p$, $\{ c(t) | 0 \leq t \leq 1, c(0)=c(1)=p\}$. Take a vector $X$ in $T_p M$ and parallel transport along a closed curve $c(t)$; we end up with a new vector $X_c \in T_p M$. Thus, the loop $c(t)$ and the connection $\nabla$ induce a linear transformation $P_c: T_pM \to T_pM$. The set of all these transformations is denoted by ${\rm Hol}_p (M)$ and called the {\it holonomy group at $p$}.  
\end{definition}

The holonomy group measures how vectors are transformed by parallel transport around a closed curve at a point $p$ of $M$. In fact, holonomy groups can be defined more generally for a vector bundle $E$ with a connection, not necessarily the tangent bundle with its affine connection; the reader is referred to \cite{Joyce:2000} for a detailed discussion of holonomy.

Note that ${\rm Hol}_p (M)$ must be a subgroup of $GL(n,\IR)$, which is the maximal holonomy group possible. ${\rm Hol}_p (M)$ is trivial if and only if the Riemann tensor vanishes.

Now, suppose that $M$ is connected (which we always assume in these lectures), and that $p$ and $q$ are two points of $M$ connected by a curve $a$. The curve $a$ defines a map $\t_a: T_pM \to T_qM$ by parallel transporting a vector in $T_pM$ to $T_qM$ along $a$. Then the holonomy groups are related by ${\rm Hol}_p (M) = \t_a^{-1} {\rm Hol}_q (M) \t_a$, hence ${\rm Hol}_q (M)$ is isomorphic to ${\rm Hol}_p (M)$. For that reason, the holonomy group ${\rm Hol}_p (M)$ is independent of the base point $p$, and we usually omit the subscript $p$ and denote by ${\rm Hol} (M)$ the {\it holonomy group of a manifold $M$}.

In particular, if $M$ is a Riemaniann manifold and $\nabla$ is a metric connection, then parallel transport preserves the length of a vector, which implies that the holonomy group must be a subgroup of $SO(n)$ (if $M$ is orientable).

Now what is the holonomy of a complex manifold with a K\"ahler metric?

\begin{proposition}\label{UNhol}
Let $M$ be a complex manifold of real dimension $2m$, with a K\"ahler metric $g$. The holonomy group of $M$ is contained in $U(m)$.
\end{proposition}

Take a vector $X \in T_p^{(1,0)}M$ in the holomorphic tangent space at a point $p$ of $M$. We saw in section \ref{kahlerdef} that for an affine connection $\nabla$ corresponding to a K\"ahler metric, vectors with holomorphic indices remain with holomorphic indices after parallel transport. Therefore, if $X_c$ is the vector resulting from parallel transporting $X$ around a closed loop $c$, we must have $X_c \in T_p^{(1,0)}M$. Moreover, $\nabla$ preserves the length of a vector. This implies that the holonomy group is the set of all transformations $P_c: T_p^{(1,0)}M \to T_p^{(1,0)}M$ which preserve the length of a vector, which is (a subgroup of) $U(m)$.

Note that we can use proposition \ref{UNhol} as a definition of K\"ahler manifolds; all the above properties of K\"ahler manifolds follow from it.

\section{\CY\ geometry}

We are now ready to investigate \CY\ manifolds, which are a particular kind of K\"ahler manifolds.

It was in 1954 that Calabi stated his conjecture \cite{Calabi:1956,Calabi:1957}, which was proved by Yau in 1976 \cite{Yau:1977,Yau:1978}. Given a compact K\"ahler manifold $M$ with $c_1=0$, the proof of the conjecture guarantees the existence of a Ricci-flat K\"ahler metric on $M$, that is a K\"ahler metric with zero Ricci form. Such a manifold is called a \CY\ manifold.

However, many different definitions of \CY\ manifolds exist in the literature; we will review some of the most common definitions and study some relations among them. We will also investigate properties of \CY\ manifolds and study in details a few examples. We will end this section by quickly describing `local' \CY\ manifolds (i.e. noncompact \CY\ manifolds), which have many applications for instance in topological strings and Gromov--Witten theory.

\CY\ manifolds have been studied extensively in the recent decades, particularly because of their importance in string theory. While the mathematical study of \CY\ manifolds has helped us understand compactifications of string theory, the study of string theory has led to fascinating insights in the geometry of \CY\ manifolds, for example the study of the \CY\ moduli space and mirror symmetry. \CY\ manifolds are thus a very good example of the fruitful interactions between mathematics and physics that have been taking place in the recent decades.

\subsection{\CY\ manifolds}

Let us first list some of the most common definitions of \CY\ manifolds. A \CY\ manifold of real dimension $2m$ is a compact K\"ahler manifold $(M,J,g)$: 
\begin{enumerate}
\item{with zero Ricci form,}
\item{with vanishing first Chern class,}
\item{with ${\rm Hol}(g) = SU(m)$ (or ${\rm Hol}(g) \subseteq SU(m)$),}
\item{with trivial canonical bundle,}
\item{that admits a globally defined and nowhere vanishing holomorphic $m$-form.}
\end{enumerate}

We now describe some of the relations between these definitions. There are many ways to understand these relations. We only describe a few relations, but many others can be found in the references at the end of these lecture notes.

\noindent{\bf [1-2]} 

First, let us show that a compact Ricci-flat K\"ahler manifold has $c_1 = 0$. We saw in (\ref{chernexplicit}) that the first Chern class of a complex vector bundle $E$
over $M$ is given by $c_1 (E) = \[ {i \over 2 \p} \Tr F \]$, where $F$ is the curvature of a connection $A$. The first Chern class of a manifold was defined in section \ref{chern} to be the first Chern class of the holomorphic tangent bundle, and in this case, the curvature two-form $F$ is simply $-iR$, where $R$ is the Ricci two-form. Therefore, given a K\"ahler manifold $M$, its first Chern class is given by
\beq
c_1(M) = \[ {1 \over 2 \p} \Tr R \].
\eeq
In other words, the Ricci form defines the first Chern class of a manifold. Hence it is clear that if a K\"ahler manifold admits a Ricci-flat metric then it has $c_1=0$. However, the converse, namely, does a K\"ahler manifold with $c_1=0$ admit a Ricci-flat metric, is much more complicated to prove. It was conjectured by Calabi that the answer is yes and that the Ricci-flat metric is unique; uniqueness was proved by Calabi, existence by Yau twenty years later. More precisely, it was proved that given a complex manifold $M$ with a K\"ahler metric $g$, a K\"ahler form $\o$ and $c_1=0$, then there exists a unique Ricci-flat metric $g'$ whose K\"ahler form $\o'$ is in the same K\"ahler class as $\o$. In other words, there is a unique Ricci-flat K\"ahler metric in each K\"ahler class of $M$. This means that the Ricci-flat K\"ahler metrics on $M$ form a smooth family of dimension $h^{1,1} (M)$, isomorphic to the K\"ahler cone of $M$. A proof of this deep theorem is given in chapter 5 of \cite{Joyce:2000}. 

Therefore, $h^{1,1} (M)$ counts the number of possible Ricci-flat K\"ahler forms on a \CY\ manifold $M$. These are sometimes called the {\it K\"ahler parameters} of $M$. Moreover, as any element in a class in $H^{1,1}_{\bar{\6}} (M)$ can be used to deform the Ricci-flat K\"ahler form slightly while preserving K\"ahlerity, we can also say that $h^{1,1}(M)$ classifies infinitesimal K\"ahler deformations of the metric.

\noindent{\bf [4-5]}

We saw that the canonical bundle of a complex manifold $M$ of real dimension $2m$ is the complex vector bundle $K_M = \L^{m,0} M$, that is its sections are $(m,0)$-forms. Triviality of this bundle implies that the total space of $K_M$ is given by $M \times \IC$ (since it is a line bundle). Therefore, corresponding to the unit section $M \times \{1\}$, that is the constant function $1$, there is a globally defined and nowhere vanishing holomorphic $(m,0)$-form $\O$, which is usually called the {\it holomorphic volume form}. Moreover, it is clear that any globally defined $(m,0)$-form can be written as $f \O$ for some function $f$ on $M$. But, if $M$ is compact and the form is holomorphic, $f$ must be holomorphic, and the extension of the maximum modulus principle of complex analysis tells us that $f$ is constant. Hence $h^{m,0} = 1$. On the other hand, the existence of a globally defined and nowhere vanishing holomorphic $(m,0)$-form $\a$ directly implies that the canonical bundle is trivial.

\noindent{\bf [2-4]}

The canonical bundle is the determinant line bundle of the holomorphic cotangent bundle, i.e. it is the highest antisymmetric tensor product of the holomorphic cotangent bundle. We saw in section \ref{chern} that $c_1(E)=0$ is the same thing as the determinant line bundle $\L^k E$ being trivial, where $k$ is the rank of $E$. Therefore $K_M = \L^m T^{*(1,0)}M$ is trivial if and only if $c_1(T^{*(1,0)} M) = -c_1(T^{(1,0)}M)=-c_1 = 0$. 

\noindent{\bf [1-3]}

We now show that if $M$ is a Ricci-flat K\"ahler manifold of real dimension $2m$, then its holonomy group is contained in $SU(m)$. Let $V = V^{k} \6_{k} \in T_p M$ be a tangent vector, and parallel transport it along an infinitesimal parallelogram of area $\d a^{mn}$ with edges that are parallel to the vectors $\6_m$ and $\6_n$. It is a standard result that
\beq
V'^k = V^k + \d a^{mn} {{R_{mn}}^k}_l V^l.
\eeq
The matrices $\d^k_l + \d a^{mn}{{R_{mn}}^k}_l$ are the elements of the holonomy group that are infinitesimally close to the identity. For a K\"ahler metric, the matrices $\d a^{mn} {{R_{mn}}^k}_l$ are in the Lie algebra of $U(m)$. In fact, in a neighborhood of the identity we have that $U(m) \cong SU(m) \times U(1)$, where the $U(1)$ factor is generated by the trace $\d a^{mn} {{R_{mn}}^k}_k$. One can easily show that this is equal to $-4 \d a^{\m \bar{\n}} R_{\m \bar{\n}}$. Therefore, if the metric is Ricci-flat, then the $U(1)$ part of the holonomy vanishes, and the holonomy groups must be contained in $SU(m)$. The converse is also true; if the holonomy of a K\"ahler manifold is contained in $SU(m)$, then its K\"ahler metric is Ricci-flat.

In fact, we have only shown that this is true for simply connected manifolds, that is for closed curves that can be continuously shrunk to a point. The equivalence still holds for multiply connected manifolds, but the proof is more involved.

\subsection{Cohomology}

The Hodge numbers of a \CY\ manifold satisfy a few more properties, which drastically decrease the number of undetermined Hodge numbers. We will now focus on \CY\ threefolds, that is \CY\ manifolds with complex dimension $3$, for the sake of brevity. These are the most important \CY\ manifolds in string theory applications. But most results extend straighforwardly to higher dimensional \CY\ manifolds.

We have already shown that the Hodge numbers of K\"ahler manifolds satisfy a Hodge star duality $h^{p,q} = h^{3-q,3-p}$ and a complex conjugation duality $h^{p,q}=h^{q,p}$. For \CY\ manifolds, there is a further duality, sometimes called {\it holomorphic duality}. We have shown in the previous section that triviality of the canonical bundle of a \CY\ manifold $M$ of real dimension $6$ implies that $h^{3,0} = 1$, i.e. the existence of a unique holomorphic volume form $\O$. Given a $(0,q)$ cohomology class $[\a]$, there is a unique $(0,3-q)$ cohomology class $[\b]$ such that $\int_M \a \wedge \b \wedge \O = 1$ (using Stoke's theorem). Thus $h^{0,q} = h^{0,3-q}$. Therefore, for a \CY\ manifold we have that $h^{3,0}=h^{0,3}=h^{0,0}=h^{3,3}=1$.

Moreover, one can show that $h^{1,0}=0$ \cite{Joyce:2000,Hubsch:1992}. Thus, $h^{1,0}=h^{0,1}=h^{0,2}=h^{2,0}=h^{2,3}=h^{3,2}=h^{3,1}=h^{1,3}=0$. Therefore, the only remaining independent Hodge numbers are $h^{1,1}$ and $h^{2,1}$, and the Hodge diamond takes the form:
\beq
\begin{array}{ccccccc}
&&&1&&& \\
&&0&&0&& \\
&0&&h^{1,1}&&0&\\
1&&h^{2,1}&&h^{2,1}&&1\\
&0&&h^{1,1}&&0&\\
&&0&&0&& \\
&&&1&&& \\
\end{array}
\eeq

The Euler characteristic of a \CY\ manifold accordingly simplifies. Recall that $\c = \sum_{k=0}^{2m} (-1)^k b^k$, so we now have that $\c=2b^0-2b^1+2b^2-b^3 = 2-0+2h^{1,1}-2-2h^{2,1}$, that is
\beq
\c = 2(h^{1,1}-h^{2,1}).
\eeq

Therefore, if the Euler characteristic is easily computed, we only have to compute one of the two independent Hodge numbers to get all the topological information. In fact, we saw in section \ref{chern} that the Euler characteristic is given by the integral over $M$ of the top Chern class of $M$, which is $c_3(M)$ for a \CY\ threefold:
\beq
\c = \int_M c_3(M).
\eeq
This formula can be used to compute the Euler characteristic of $M$. 

We saw earlier that $h^{1,1}$ classifies infinitesimal deformations of the K\"ahler structure. For a \CY\ threefold, similarly, $h^{2,1}$ classifies infinitesimal deformations of the complex structure. We refer the reader to chapter 6 of \cite{Hori:2003} for a detailed discussion of this interpretation and of the moduli space of \CY\ manifolds.

\begin{remark}
One of the fascinating property of \CY\ threefolds is that they come in mirror pairs, $(M,W)$, such that $H^{2,1} (W) \cong H^{1,1} (M)$ and $H^{1,1} (W) \cong H^{2,1} (M)$. Roughly speaking, the complex structure moduli is exchanged with the K\"ahler structure moduli. This is the basic idea behind {\it mirror symmetry}. See \cite{Hori:2003,Cox:1999} for more information on this subject.
\end{remark}

\subsection{Examples}

We will now study in some details two particular examples of \CY\ threefolds: the quintic in $\IC \IP^4$, and the Tian-Yau manifold. Both examples have been very important in the history of string theory, and they will help us find our way in the asbtract jungle of \CY\ threefolds.

There are various ways one can follow to see if a K\"ahler manifold is \CY. The more `hands-on' way is probably to find a globally defined and nowhere vanishing holomorphic volume form (see for instance chapter 9 of \cite{Candelas:1987is} for this approach). Another approach, more abstract, is to compute explicitely the first Chern class of the manifolds and see that it vanishes. In our two examples, we will first follow the latter, as in the process we will learn how to compute Chern classes. Then we will quickly review how to construct the holomorphic volume form.

Through these two examples we will study in more generality complete intersection manifolds in complex projective spaces and products thereof. But to start with we need to know the Chern classes of the complex projective spaces $\IC \IP^m$.

Obviously we cannot prove here all the results that are needed to carry on the computations. The reader is referred to \cite{Hubsch:1992} for a detailed discussion of various constructions of \CY\ threefolds. 

\subsubsection{Chern classes of $\IC \IP^m$}

First, we need to compute the total Chern class of $\IC \IP^m$. We recall that homogeneous coordinates $z_i$, $i=0,\ldots,m$ of $\IC^{m+1}$ are sections of the hyperplane line bundle $L$. Thus, the holomorphic tangent bundle of $\IC^{m+1}$ is spanned by tangent vectors $s_i (z) {\6 \over \6 z_i}$, where the $s_i$ are any sections of $L$. Now, on $\IC \IP^m$, the holomorphic tangent bundle $T^{(1,0)} \IC \IP^m$ is also spanned by $s_i (z) {\6 \over \6 z_i}$, with the $s_i$ any sections of the hyperplane line bundle --- which we now denote by $\co_{\IC \IP^m} (1)$, but we have to take equivalence classes with respect to overall rescaling, since overall rescaling is trivial in $\IC \IP^m$. That is, we have a map from $\co_{\IC \IP^m} (1)^{\bigoplus(m+1)}$ to $T^{(1,0)} \IC \IP^m$ such that its kernel is the trivial line bundle $\IC$ of multiples of a nowhere-vanishing generator $(z_0,\ldots,z_m) \mapsto z_i {\6 \over \6 z_i} \cong 0$ in $\IC \IP^m$. This is summarized in the following exact sequence, called the {\it Euler sequence}:
\beq
0 \to \IC \to \co_{\IC \IP^m} (1)^{\bigoplus(m+1)}\to T^{(1,0)} \IC \IP^m \to 0.
\eeq
Trivially $c(\IC)=1$, so by properties of Chern classes we have that $c(\IC \IP^m)= c(T^{(1,0)} \IC \IP^m) = c (\co_{\IC \IP^m} (1)^{\bigoplus(m+1)}) = \[c (\co_{\IC \IP^m} (1)\]^{m+1}$. $\co_{\IC \IP^m} (1)$ is a line bundle, that is its fibers are one-dimensional, and so the expansion of the total Chern class is simply $c(\co_{\IC \IP^m} (1)) = 1 +c_1(\co_{\IC \IP^m} (1))$. If we let $x = c_1(\co_{\IC \IP^m} (1))$, we find that
\beq
c(\IC \IP^m)=(1+x)^{m+1}.
\eeq

\subsubsection{\CY\ condition for complete intersection manifolds}

We now want to see what the \CY\ condition is for complete intersection manifolds. But let us first look at Calabi-Yau hypersurfaces in complex projective spaces.

\begin{figure}[htp]
\begin{center}
\psfrag{CP2}{$\IC \IP^2$}
\psfrag{X}{$X$}
\psfrag{TX}{$T^{(1,0)} X$}
\psfrag{NX}{$N_X$}
\includegraphics[width=6cm]{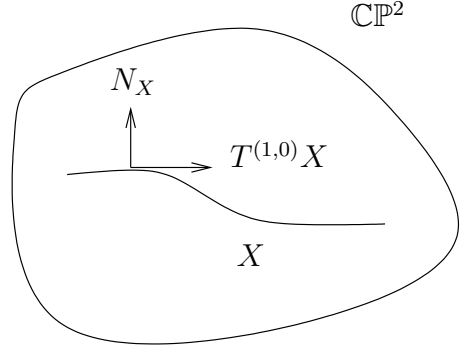}
\caption{Representation of the normal bundle of a one-dimensional hypersurface $X$ in $\IC \IP^2$, which can be understood as the quotient $N_X = \frac{T^{(1,0)}\IC \IP^2|_X}{T^{(1,0)}X}$.}
\label{f:normal}
\end{center}
\end{figure}

Let $X$ be a smooth hypersurface in $\IC \IP^m$ defined as the zero-locus of a degree $d$ polynomial $p$. We can see $p$ as a section of the holomorphic line bundle $\co_{\IC \IP^m} (d)$. Consider $T^{(1,0)}X$, the holomorphic tangent bundle of $X$. We define the normal bundle $N_X$ on $X$ to be the quotient $N_X = \frac{T^{(1,0)}\IC \IP^m|_X}{T^{(1,0)}X}$. As a result, we have the exact sequence $0 \to T^{(1,0)}X \to T^{(1,0)}\IC \IP^m|_X \to N_X \to 0$. To convince oneself that this definition of the normal bundle makes sense, one can visualize it for a one-dimensional hypersurface in $\IC \IP^2$, as in figure \ref{f:normal}.

Now, roughly speaking, on $X$ the section $p$ maps points of $X$ to $0$ in the fibers of $\co_{\IC \IP^m} (d)$, since $X$ is defined as the zero-locus of $p$. Thus, $p$ serves as a coordinate near $X$, and in fact the normal bundle $N_X$ of $X$ is simply $\co_{\IC \IP^m} (d)|_X$. This is the crucial point in the computation. The above exact sequence then becomes (this is also known as the adjunction formula 1; see \cite{Griffiths:1978} for more about this)
\beq
0 \to T^{(1,0)}X \to T^{(1,0)}\IC \IP^m|_X \to \co_{\IC \IP^m} (d)|_X \to 0,
\eeq
which implies that $c(X) = c(\IC \IP^m) / c(\co_{\IC \IP^m}(d))$. $\co_{\IC \IP^m}(d)$ is a line bundle, so $c(\co_{\IC \IP^m}(d))=1+c_1(\co_{\IC \IP^m}(d))$. We know from above that  $c(\co_{\IC \IP^m} (1))=1+x$. Therefore, its Chern character is $ch(\co_{\IC \IP^m} (1))=e^x$. It follows that $ch(\co_{\IC \IP^m} (d))=e^{dx} = 1 + c_1(\co_{\IC \IP^m} (d))+\ldots$, hence
\beq
c(\co_{\IC \IP^m}(d))=1+dx.
\eeq
Using these results we find that the Chern class of $X$ is
\beq
c(X)= {(1+x)^{m+1} \over 1+dx}.
\eeq

Since $x$ is a closed two-form, we can expand $c(X)$ in such a way that products of $x$ are wedge products, from which we can extract the Chern classes $c_k(X)$ of $X$. The first Chern class is 
\beq
c_1(X) = [(m+1)-d]x.
\eeq
Therefore, we have found an explicit realization of the \CY\ condition for hypersurfaces; the condition $c_1=0$ implies a condition on the degree of the polynomial equation, $d=m+1$. If we want a \CY\ threefold, we have $m=4$, and therefore $X$ must be given by the zero-locus of a degree $5$ polynomial, that is a quintic in $\IC \IP^4$.

In fact, it is straightforward to generalize the above computation to complete intersection manifolds $Y$ given by the zero-locus of a finite number of polynomials in $\IC \IP^m$. If there are $l$ such polynomials of degree $d_i$, $i=1,\ldots,l$, the Chern class is given by
\beq
c(Y)= {(1+x)^{m+1} \over \prod_{i=1}^l (1+d_i x)}.
\eeq
Therefore, $c_1(Y)=0$ implies the restriction $m+1 = \sum_{i=1}^l d_i$ on the degrees of the $l$ polynomials defining $Y$.

There are only five solutions to the above condition, if we are looking for \CY\ threefolds, that is $l=m-3$ --- and we assume $d_i \geq 2$, since if one equation has $d_j=1$ the manifold defined by the complete intersection in $\IC \IP^m$ is equivalent to the manifold defined by the other equations in $\IC \IP^{m-1}$: the quintic in $\IC \IP^4$, the intersection of two cubics in $\IC \IP^5$, the intersection of a quadric and a quartic in $\IC \IP^5$, the intersection of two quadrics and a cubic in $\IC \IP^6$, and the intersection of four quadrics in $\IC \IP^7$.

Finally, if we expand completely the total Chern class (using the fact that $c_1=0$), we find
\beq
c(Y)=1 + \2 \[(\sum_{i=1}^l d_i^2)-(n+1)\]x^2 -{1\over 3} \[(\sum_{i=1}^l d_i^3)-(n+1)\]x^3.
\eeq
This result will be useful for the computation of the Euler characteristic through the integration of the third Chern class over the manifold $Y$.

\subsubsection{The quintic in $\IC \IP^4$}\label{quintic}

We will now concentrate on the quintic $Q$ in $\IC \IP^4$, which is given by a polynomial equation of degree $5$ in the homogeneous coordinates of $\IC \IP^4$. According to the results of the previous section, we have that
\beq
c(Q) = 1 + 10 x^2 - 40 x^3. 
\eeq
To find its Euler characteristic, we must integrate $c_3 = -40x^3$ over $Q$. How do we do that? We invoke Poincar\'e duality, which is an intersection pairing of cohomology classes.\footnote{Let $M$ be a $n$-dimensional manifold. Using the operators defined in section 2, one can show that a $k$-form $\o$ on $M$ is harmonic if and only if $*\o$ is also harmonic. Since the space of harmonic $k$-forms is isomorphic to the de Rham cohomology group $H_{dR}^k (M, \IR)$, and that $*\o$ is a $(n-k)$-form, there is an isomorphism between $H_{dR}^k (M, \IR) \cong H_{dR}^{n-k} (M, \IR)$. This is Poincar\'e duality in a nutshell.} In fact, we want to `lift' the integral to the embedding space where the integration is trivial. Using Poincar\'e duality and de Rham's theorems relating homology and cohomology one can prove the following theorem about complex integration over submanifolds: \cite{Hori:2003,Candelas:1987is}

\begin{theorem}
Let $X$ be a closed $k$-dimensional submanifold of $M$, where $M$ is $n$-dimensional. For any closed form $\t \in H^{k}_{dR} (M,\IR)$, we can define the integration $\int_X \t$. By Stoke's theorem, this integral is independent of the choice of representative of the cohomology class. Thus $\int_X$ is a linear map $H^k \to \IR$, and Poincar\'e duality says that there is a $(n-k)$-form $\h_X \in H^{n-k}_{dR} (M,\IR)$ such that
\beq
\int_X \t = \int_M \t \wedge \h_X.
\eeq
We call $\h_X$ the {\rm Poincar\'e dual class to $X$}.
\end{theorem}

Therefore, the Poincar\'e dual class restricts the integration to the submanifold $X$ like a delta function. But how do we find $\h_X$? In fact in the special case where the normal bundle $N_X$ to $X$ is the restriction to $X$ of some bundle over $M$, i.e. $N_X = E |_X$ --- which is the case we consider, with $E = \co_{\IC \IP^4} (5)$ --- then it is easy to find $\h_X$;  $\h_X = c_k (E)$, where $k$ is the rank of $E$; that it is it is the top Chern class of $E$.

Thus, for the quintic we have that $\h_Q = c_1 (\co_{\IC \IP^4} (5)) = 5x$. Now, since $\int_{\IC \IP^m} x^m = 1$ (this is so because $x$ is Poincar\'e dual to a hyperplane and $m$ hyperplanes intersect at a point), we find that
\beq
\c(Q) = \int_Q c_3(Q) = \int_Q (-40 x^3) = \int_{\IC \IP^4} (-40x^3) \wedge (5x)=-200.
\eeq

Now to pursue the study of the quintic further we must determine its Hodge numbers. Let us first consider $h^{2,1}$, which classifies infinitesimal deformations of the complex structure. In other words, given a polynomial equation of a certain degree in $\IC \IP^m$, the complex structure is determined by the free coefficients (usually called parameters) in the polynomial equation. Fixing these parameters `chooses' a particular complex structure; but by modifying these coefficients we move in the moduli space of complex structures of the polynomial equation of a certain degree. Therefore, these parameters classify infinitesimal deformations of the complex structure, and $h^{2,1}$ of the complete intersection manifold is equal to the number of free parameters.

\begin{remark}
Note that this simple method for finding the Hodge numbers of a \CY\ manifold does not always work; the manifold must satisfy a few extra conditions. However it works in this example and in the next example we will consider. Another method is to use the Lefschetz hyperplane theorem to compute $h^{1,1}$ directly --- see \cite{Hubsch:1992} for an explanation of this technique.
\end{remark}

For the quintic in $\IC \IP^4$, there are initially $126$ parameters.\footnote{The number of independent degree $d$ homogeneous polynomials in $n$ variables is given by the binomial coefficient $\begin{pmatrix}d+n-1 \\ n-1 \end{pmatrix}$.} The group of holomorphic automorphisms of $\IC \IP^m$ being $PGL(m+1,\IC)$,\footnote{The projective linear group $PGL(m+1,\IC)$ is the general linear group $GL(m+1,\IC)$ quotiented by the group $Z(m+1)$ of all nonzero scalar transformations, that is $PGL(m+1,\IC) = GL(m+1,\IC) / Z(m+1)$. This is the group of holomorphic automorphisms of $\IC \IP^m$ since the action of $GL(m+1,\IC)$ on $\IC^{m+1}$ descends to an action of $PGL(m+1,\IC)$ on $\IC \IP^m$.} $25-1$ of them can be removed by an homogeneous linear change of variables. Moreover, one parameter corresponds to overall rescaling. Hence, there are $h^{2,1} (Q) = 126 - (25-1) -1 = 101$ parameters describing the complex structure of $Q$.

Since $\c = 2(h^{1,1}-h^{2,1})$, we have that $h^{1,1} (Q) =1$, i.e. there is only one Ricci-flat K\"ahler form on the quintic. Hence the K\"ahler structure moduli space is one-dimensional.

To summarize our result; the quintic $Q$ in $\IC \IP^4$ has Euler class $\c = -200$ and Hodge diamond
\beq
\begin{array}{ccccccc}
&&&1&&& \\
&&0&&0&& \\
&0&&1&&0&\\
1&&101&&101&&1\\
&0&&1&&0&\\
&&0&&0&& \\
&&&1&&& \\
\end{array}
\eeq

Now we know that $Q$ is a \CY\ manifold, and we studied it using Chern classes. We will now construct a holomorphic volume form $\O$ on $Q$. Alternatively, we could have started our study of the quintic by attempting a direct construction of an holomorphic volume form, and show that way that $Q$ is indeed \CY.

Define the form $\t$ on $\IC^5$ by $\t = \sum_{\m=0}^4 dz_{0} \wedge \ldots \wedge z_{\m} \wedge \ldots \wedge dz_{4}$ (notice that we have replaced $dz_{\m}$ by $z_{\m}$). $\t$ is clearly a holomorphic $(4,0)$-form. However, it is not invariant under scaling $z_{\m} \to \l z_{\m}$, so it is not well-defined on $\IC \IP^4$. But the form $\t / Q$ is invariant, where $Q$ is a degree $5$ homogeneous polynomial in $\IC \IP^4$. However, it is singular at $Q=0$.

Now let $\g_Q$ be a small loop around $Q=0$ in $\IC \IP^4$. Define
\beq
\O = \int_{\g_Q} {\t \over Q}.
\eeq
$\O$ is a globally defined and nowhere vanishing holomorphic $(3,0)$-form on $Q=0$. To see this, in a coordinate patch, rewrite $dz_0 = \({\6 z_0 \over \6 Q}\) dQ$, and integrate along the loop $\g_Q$; by the residue theorem we find
\beq
\O = (2 \p i) \({\sum_{\mu=1}^4 dz_1 \wedge \ldots \wedge z_\mu \wedge \ldots \wedge dz_4 \over (\6 Q / \6 z_0 )}\)_{Q=0},
\eeq
which is a nowhere vanishing holomorphic $(3,0)$-form on $Q=0$.

We have found an holomorphic volume form $\O$ on $Q$; therefore as we saw in the previous section all other holomorphic $(3,0)$-forms are constant multiples of $\O$.

We can easily extend this construction to complete intersection manifolds constructed as the zero-locus of a finite number of polynomials in a projective space.

Consider the complete intersection of $N$ polynomials $P^i$, $i=1,\ldots,m-3$ in a projective space $\IC \IP^{m}$ (in order to have a threefold). Define the form $\t = \sum_{\m=0}^{m} dz_{0} \wedge \ldots \wedge z_{\m} \wedge \ldots \wedge dz_{m}$ on $\IC^{m+1}$. Again, this is not invariant under scaling. However, define the form $\t / (\prod_{i=1}^{m-3} P^i)$; it is invariant under scaling if the degrees $d_i$ of the polynomials $P^i$ satisfy
\beq
(m+1) =\sum_{i=1}^{m-3} d_i ,
\eeq
which is exactly the restriction on the degrees of the polynomial that we found earlier for the first Chern class to be zero.

Now, consider a $(m-3)$-dimensional contour
\beq
\G_{m-3} = \g_1 \times \g_2 \times \ldots \times \g_{m-3},
\eeq
which is the Cartesian product of $(m-3)$ small loops around the $(m-3)$ curves defined by $P^i=0$, $i=1,\ldots,m-3$. Define the form
\beq
\O = \int_{\G_{m-3}} {\t \over \prod_{i=1}^{m-3} P^i};
\eeq
this is a globally defined and nowhere vanishing holomorphic $(3,0)$-form on the complete intersection of the polynomials $P^i$ in $\IC \IP^m$.

\subsubsection{The Tian-Yau manifold}\label{CICY}

In order to study the Tian-Yau manifold, we must generalize the results of the last section to manifolds defined by the zero-locus of a finite number of homogeneous polynomial equations in a product of projective spaces. Let us first introduce some notation.

Such spaces will be denoted by a {\it configuration matrix} which gives the degree of each polynomial in the variables of each projective space. Each column corresponds to the degree of one of the polynomial. We also usually add the Euler characteristic of the manifold at the bottom right of the configuration matrix. For instance, in this notation the quintic in $\IC \IP^4$ is given by
\beq
\IC \IP^4 | 5 |_{-200},
\eeq
while the Tian-Yau manifold is given by
\beq
\begin{array}{c|ccc|}
\IC \IP^3 & 1 & 3 & 0 \nonumber\\
\IC \IP^3 & 1 & 0 & 3
\end{array}_{-18}.
\eeq
In other words, the Tian-Yau manifold is given by three polynomial equations in $\IC \IP^3 \times \IC \IP^3$; one of degree $1$ in both $\IC \IP^3$, one of degree $3$ in the first $\IC \IP^3$, and one of degree $3$ in the second $\IC \IP^3$. That is, if $x_{\m}$ and $y_{m}$ are respectively homogeneous coordinates of the two $\IC \IP^3$, it represents the system of equations
\beq
f^{\m \n \r} x_{\m} x_{\n} x_{\r} =0,~~~~g^{m n r} y_m y_n y_r=0,~~~~h^{\m m} x_{\m} y_m =0,
\eeq
where $f,g,h$ are coefficients of the equations.

\begin{remark}
In fact, the configuration matrix does not specify a particular manifold, but rather the family of all complete intersections parameterized by the space of coefficients; we call this family a {\it configuration}. As we noted earlier, two different sets of coefficients correspond to two complete intersections which in general are different as complex manifolds. The space of these coefficients is a parameter space for the complete intersection manifolds, and by taking into account automorphisms of the ambient space and overall rescaling it parameterizes the complex structure moduli space of this family of complete intersections.
\end{remark}

Therefore, we should say that the Tian-Yau manifold is an element of the configuration above, rather than the configuration itself. In fact, it is given by the following equations with fixed coefficients:
\beq
\sum_{i=0}^3 x_i y_i=0,~~~~\sum_{i=0}^3 (x_i)^3=0,~~~\sum_{i=0}^3 (y_i)^3=0.
\eeq

Now we want to compute the Euler characteristic and the Hodge numbers of this configuration. It is straightforward to generalize the results of the previous section. 

Let $X$ be a smooth complete intersection manifold defined by the configuration matrix
\beq
\begin{array}{c|ccc|}
\IC \IP^{n_1} & d_1^1 &\cdots&d_N^1 \nonumber\\
\vdots & \vdots &&\vdots \nonumber\\
\IC \IP^{n_l} & d_1^l &\cdots&d_N^l
\end{array},
\eeq
that is it is a complete intersection manifold in a product of $l$ projective spaces of dimensions $n_i$, $i=1,\ldots,l$, defined by the zero-locus of $N$ polynomials of degree vectors ${\bf d_j}$, $j=1,\ldots,N$ in the $l$ projective spaces. Given such a configuration, we can generalize the previous computation of the Chern class to obtain
\beq
c(X) = {\prod_{r=1}^{l} (1+x_r)^{n_r+1} \over \prod_{a=1}^N (1 + \sum_{s=1}^l d_a^s x_s)}.
\eeq
By expanding, the first Chern class is
\beq
c_1 (X) = \sum_{r+1}^{l} \(n_r +1 - \sum_{a=1}^N d_a^r\) x_r.
\eeq
For $c_1(X)$ to be zero, all the coefficients in the sum must vanish, and we find the condition
\beq\label{CYcondTY}
\sum_{a=1}^N d_a^r =n_r +1,~~~~~ \forall~r=1,\ldots,l. 
\eeq

For the Tian-Yau manifold, $l=2$, $n_1=n_2=3$, $N=3$ and $d_1^1=d_1^2=1$, $d_2^1=3,d_2^2=0$ and $d_3^1=0,d_3^2=3$. The condition is satisfied, and therefore the Tian-Yau manifold is a \CY\ manifold.

To find its Euler characteristic, we must expand the total Chern class to find an expression for the third Chern class. If $c_1(X)=0$, we find that
\beq
c_3(X) = \sum_{r,s,t=1}^l \({1 \over 3} \[ \d^{rst}(n_r+1) - \sum_{a=1}^N d_a^r d_a^s d_a^t\] \) x_r x_s x_t.
\eeq
To integrate this result, we need a Poincar\'e dual class. As before, the normal bundle to $X$ is restriction of a bundle on the covering space $M$, that it $N_X = E |_X$ for some bundle $E$ over $M$. In fact, $E = \bigoplus_{a=1}^N \( \bigotimes_{r=1}^l \co_r(d_a^r) \)$. Therefore, $E$ is a bundle of rank $N$, hence $\h_X=c_N(E)$. Since 
\beq
c\[\bigoplus_{a=1}^N \( \bigotimes_{r=1}^l \co_r(d_a^r) \)\]=\bigwedge_{a=1}^N c\( \bigotimes_{r=1}^l \co_r(d_a^r) \),
\eeq
and $\bigotimes_{r=1}^l \co_r(d_a^r)$ is a line bundle for any $a$, we then have that
\beq
\h_X = c_N\[\bigoplus_{a=1}^N \( \bigotimes_{r=1}^l \co_r(d_a^r) \)\] = \bigwedge_{a=1}^N c_1\( \bigotimes_{r=1}^l \co_r(d_a^r) \) =\bigwedge_{a=1}^N \(\sum_{r=1}^l d_a^r x_r\).
\eeq
Thus, since $\int_{\IC \IP^{n_i}} (x_i)^{n_i} =1$ for $i=1,\ldots,l$, we find that the result of the integral is the coefficient of $\L_{r=1}^l (x_r)^{n_r}$, the volume form on $X$ --- which we denote by the subscript `top' --- in the following expression:
\beq
\c(X) = \[ \sum_{r,s,t=1}^l \({1 \over 3} \[ \d^{rst}(n_r+1) - \sum_{a=1}^N d_a^r d_a^s d_a^t\] \) x_r x_s x_t \cdot \bigwedge_{b=1}^N \(\sum_{p=1}^l d_b^p x_p\) \]_{\rm top}.
\eeq
Using this general formula one can compute easily that the Euler characteristic of the Tian-Yau manifold is $\c = -18$. Furthermore, one can show that the Euler characteristic of any complete intersection \CY\ manifold must be nonpositive, that is $\c \leq 0$.

Now let us try to find the Hodge numbers of the Tian-Yau manifold. We will use the same method as for the quintic, namely simply counting the free parameters in the polynomial equations.

Two equations are of degree $3$ in $4$ variables, so together they have $40$ free parameters. However, in each $\IC \IP^3$ $(16-1)$ of them can be removed by a homogeneous linear change of variables, and $1$ by overall rescaling. Therefore in these two equations there are in total $8$ free parameters.

Now the third equation has $16$ coefficients, and $1$ can be removed by oversall rescaling. 

Therefore, in total there are $15+8=23$ free parameters. Hence $h^{2,1} (X) = 23$. Further, from the equation $\c = 2 (h^{1,1}-h^{2,1})$, we find that the Tian-Yau manifold has Euler characteristic $\c= -18$ and Hodge diamond
\beq
\begin{array}{ccccccc}
&&&1&&& \\
&&0&&0&& \\
&0&&14&&0&\\
1&&23&&23&&1\\
&0&&14&&0&\\
&&0&&0&& \\
&&&1&&& \\
\end{array}
\eeq

We can construct the holomorphic volume form in exactly the same way as we did before. Let $X$ be a smooth manifold given by the configuration matrix
\beq
\begin{array}{c|ccc|}
\IC \IP^{n_1} & d_1^1 &\cdots&d_N^1 \nonumber\\
\vdots & \vdots &&\vdots \nonumber\\
\IC \IP^{n_l} & d_1^l &\cdots&d_N^l
\end{array},
\eeq
Let $z^r_i$, $i=0,\ldots,n_r$ be coordinates on $\IC^{n_r+1}$. On each complex space we define a form $\t_r = \sum_{\m=0}^{m+1} dz^r_{0} \wedge \ldots \wedge z^r_{\m} \wedge \ldots \wedge dz^r_{m}$. The product of all these, $\t = \prod_{r=1}^l \t_r$, is a form on the space $\prod_{r=1}^l \IC^{n_r+1}$. Again, this is not invariant under scaling. However, define the form $\t / (\prod_{a=1}^{N} P^a)$; it is invariant under scaling if the condition (\ref{CYcondTY}) is satisfied, and thus defined on the space $\prod_{r=1}^l \IC \IP^{n_r}$.

Now, consider a contour
\beq
\G_{N} = \g_1 \times \g_2 \times \ldots \times \g_{N},
\eeq
which is the Cartesian product of $N$ small loops around the $N$ curves defined by $P^i=0$, $i=1,\ldots,N$. Define the form
\beq
\O = \int_{\G_{N}} {\t \over \prod_{a=1}^{N} P^a};
\eeq
this is a globally defined and nowhere vanishing holomorphic $(3,0)$-form on the complete interesection of the polynomials $P^N$ in $\prod_{r=1}^l \IC \IP^{n_r}$.

The Tian-Yau manifold was important historically as it was the first manifold to yield a three-generation spectrum for the low-energy physics coming out of string theory. The first attempts at finding the standard model from string theory used what is now called the `standard' compactification of the $E_8 \times E_8$ heterotic string theory. In these compactifications, the number of generations of the low-energy theory is given by half the absolute value of the Euler characteristic of the compact \CY\ threefold. Therefore, to find a three-generation model we must have $\c = \pm 6$.

The Tian-Yau manifold does not satisfy this condition; however, it admits a free $\IZ_3$ action which leads to a new (non-simply connected) \CY\ threefold, $X / \IZ_3$, which has Euler characteristic $\c = -18/3 = -6$. Therefore, compactification of heterotic strings on the quotient threefold yields a three-generation model.

It may seem easy to construct three-generation manifolds, since there is a large number of complete intersection \CY\ threefolds in products of projective spaces (at least a few thousands). But in fact, only a few phenomelogically interesting constructions have been found that way, and they all seem to be simply related (see \cite{Hubsch:1992}). This is rather surprising, especially because Tian and Yau constructed their manifold before a list of complete intersection \CY\ threefolds was even compiled.

However, at the moment there are many other ways to construct \CY\ manifolds, although the technique we have explained is still probably the simplest one. For instance, one can construct \CY\ manifolds as hypersurfaces or complete intersections in weighted projective spaces, as blow-up of orbifolds, as double fibrations, as hypersurfaces or complete intersections in toric manifolds, etc. These constructions yield many other three-generation manifolds.

Moreover, the standard compactification of heterotic strings was the first attempt at extracting real physics from string theory, but there are now many other ways to obtain phenomenologically interesting physics from string theory. For example, one can work in type II theory, or consider `non-standard' compactifications of heterotic strings, that is more complicated compactifications where a vector bundle which is not the tangent bundle is embedded in the visible $E_8$ gauge group of the $E_8 \times E_8$ heterotic string. The last approach has been pursued in recent years by various people (including me; this is the place where I plug some of my work rather shamelessly :-), and has led, among other things, to an interesting compactification of the $E_8 \times E_8$ heterotic string \cite{Bouchard:2005ag,Bouchard:2006} which reproduces precisely the massless spectrum of the Minimal Supersymmetric Standard Model. It would be fascinating to understand whether all these semi-realistic compactifications of string theory are somewhat related, and why should string theory have chosen one of these vacua rather than all the other non--realistic vacua floating around...

\subsection{`Local' \CY\ manifolds}

To end this section, we give a quick definition of `local' \CY\ manifolds. So far, we only considered {\it compact} manifolds, and our definitions of \CY\ manifolds assumed that the manifolds were compact. However, it is possible to generalize this definition to admit {\it noncompact} \CY\ manifolds. By local (or noncompact) \CY\ manifolds, we mean that they are open neighborhoods in compact \CY\ manifolds. These are very useful in many applications in physics, for instance in topological strings \cite{Marino:2004uf,Neitzke:2004ni}. They are also relevant in the study of geometric transitions \cite{Rossi:2004}.

Some of the relations between the various definitions of \CY\ manifolds introduced earlier become somehow tricky for noncompact \CY\ manifolds; rather than exploring these details, we will simply adopt the following definition of local \CY\ manifolds.

\begin{definition}
A {\rm \CY\ manifold} is a K\"ahler manifold $(M,J,g)$ with trivial canonical bundle.
\end{definition}

This definition applies for both compact and noncompact manifolds. The simplest noncompact \CY\ manifold is obviously $\IC^m$.

\section{Toric geometry}\label{toric}

So far we explored various aspects of complex geometry using tools of differential geometry, sometimes bifurcating in the realm of algebraic geometry. We will now focus on a subset of complex geometry, which is called toric geometry.

Toric varieties\footnote{Roughly speaking, a variety is the algebraic analog of a manifold in differential geometry. More precisely, an algebraic variety $V \subset \IC \IP^m$ is the zero locus in $\IC \IP^m$ of a collection of homogeneous polynomials. In fact, any analytic subvariety of $\IC \IP^m$ is an algebraic variety; this is a restatement of Chow's theorem mentioned in section \ref{complex}. For more information about analytic and algebraic varieties see \cite{Griffiths:1978}. In what follows we will use the terms varieties and manifolds interchangeably.} are a special kind of varieties which provide an elementary way to understand many abstract concepts of algebraic geometry. Owing to its beauty and simplicity, toric geometry also gives the possibility to compute various non-trivial results in string theory that could not be calculated otherwise.

In this section we explore various aspects of toric geometry relevant for applications in physics, mainly in string theory. Our main goal will be to construct \CY\ manifolds in toric geometry. Therefore we will skip some important concepts and applications of toric geometry. For good and more complete introductions to toric geometry, the reader is referred to \cite{Hori:2003,Skarke:1998yk,Greene:1996cy,Fulton:1993}. This section, although based on the concepts of complex geometry developed in the first three lectures, is almost independent from the rest of these lecture notes.

\subsection{Homogeneous coordinates}\label{toricCY}

Toric varieties may be approached from various points of view. They can be described using fans and homogeneous coordinates, or viewed as symplectic manifolds, or correspondingly as the Higgs branch of the space of supersymmetric ground states of the gauged linear sigma model, or even associated to convex polytopes in integral lattices. Perhaps the simplest approach is the homogeneous coordinate description \cite{Cox:1993fz}; therefore we will proceed as far as possible using this approach. 

An interesting aspect of Cox's approach to toric geometry is that by using the homogeneous coordinate construction, toric varieties look very much like the usual complex (weighted) projective spaces. In fact, from that point of view we can understand toric varieties as an algebraic generalization of (weighted) complex projective spaces.

Recall first the definition of the projective space $\IC \IP^2$. In section 1, we described $\IC \IP^2$ by embedding it into $\IC^3$:
\beq
\IC \IP^2 = (\IC^3 \setminus \{0\}) / (\IC^*),
\eeq
where the quotient is implemented by modding out by the equivalence relation
\beq
(x,y,z) \sim \l (x,y,z),
\eeq
where $\l \in \IC^*$.

Then, we generalized this definition of projective spaces by assigning weights to the coordinates of $\IC^m$. For instance, we defined the weighted projective space $\IC \IP^{(2,3,1)}$ by embedding it again into $\IC^3$:
\beq
\IC \IP^2 = (\IC^3 \setminus \{0\}) / (\IC^*),
\eeq
where the $\IC^*$ quotient is now implemented by modding out by the equivalence relation
\beq
(x,y,z) \sim  (\l^2 x,\l^3 y,\l z),
\eeq
where $\l \in \IC^*$.

Now, toric varieties may be understood as a further generalization of weighted projective spaces, where we quotient by more than one $\IC^*$ actions. That is, consider $\IC^m$ and an action by an algebraic torus $(\IC^*)^p$, $p < m$. We identify and then substract a subset $U$ that is fixed by a continuous subgroup of $(\IC^*)^p$, then safely quotient by this action to form
\beq\label{toricdef}
\cm = \( \IC^m \setminus U \) / (\IC^*)^p.
\eeq
$\cm$ is called a {\it toric variety}, as it still has an algebraic torus action by the group $(\IC^*)^{m-p}$ descending from the natural action of $(\IC^*)^m$ on $\IC^m$.

For instance, both $\IC \IP^2$ and $\IC \IP^{(2,3,1)}$ are toric varieties, as are all projective spaces and weigthed projective spaces. But there are many more toric varieties then that. 

\subsubsection{Cones and fans}

We now explain how the toric varieties introduced above can be described combinatorially using lattices. More precisely, we describe how to extract toric varieties from a fan using the homogeneous coordinate approach developed by Cox \cite{Cox:1993fz}.

Let $M$ and $N$ be a dual pair of lattices, viewed as subsets of vector spaces $M_\IR=M\otimes_\IZ\IR$ and $N_\IR=N\otimes_\IZ\IR$. Let $(u,v)\to \langle u,v \rangle$ denote the pairings $M\times N\to \IZ$ and $M_\IR\times N_\IR \to \IR$.

\begin{definition}
A {\it strongly convex rational polyhedral cone} $\s \in N_\IR$ is a set
\beq
s = \{a_1 v_1 + a_2 v_2 + \ldots + a_k v_k | a_i \geq 0 \}
\eeq
generated by a finite number of vectors $v_1,\ldots,v_k$ in $N$ such that $\s \cap (-\s) = \{ 0 \}$.
\end{definition}

Let us put words on this definition. Suppose that the lattice $N$ is $n$-dimensional, that is $N \cong \IZ^n$. A convex rational polyhedral cone is an $n$ or lower dimensional cone in $N_{\IR}$, with the origin of the lattice as its apex, such that it is bounded by finitely many hyperplanes (`polyhedra'), its edges are spanned by lattice vectors (`rational') and it contains no complete line (`strongly convex').

A face of a cone $\s$ is either $\s$ itself or the intersection of $\s$ with one of the hyperplanes bounding $\s$.

\begin{remark}
In the remaining of this section we will refer to convex rational polyhedral cones simply as cones.
\end{remark}

\begin{definition}
A collection $\S$ of cones in $N_\IR$ is called a {\it fan} if each face of a cone in $\S$ is also a cone in $\S$, and the intersection of two cones in $\S$ is a face of each.
\end{definition}

\begin{figure}[htp]
\begin{center}
\psfrag{v_1}{$v_1$}
\psfrag{v_2}{$v_2$}
\psfrag{v_3}{$v_3$}
\psfrag{CP2}{$\IC \IP^2$}
\psfrag{WP2}{$\IC \IP^{(2,3,1)}$}
\includegraphics[width=12cm]{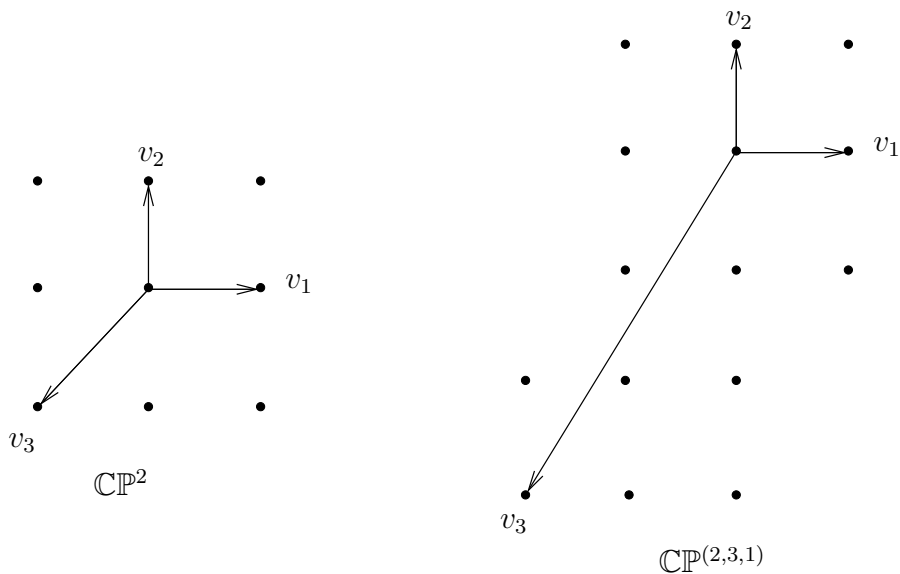}
\caption{The fan $\S$ of $\IC \IP^2$ and $\IC \IP^{(2,3,1)}$. It includes the three two-dimensional cones spanned by $v_1-v_2$, $v_2-v_3$ and $v_3-v_1$; the three one-dimensional cones $v_1$, $v_2$ and $v_3$; and the origin.}
\label{CP2}
\end{center}
\end{figure}

An example of a fan is given in figure \ref{CP2}.

Now let $\Sigma$ be a fan in $N$. Let $\Sigma (1)$ be the set of one-dimensional cones (or edges) of $\Sigma$. From now on we will focus on three-dimensional toric varieties, or correspondingly on three-dimensional lattices $M,N \simeq \IZ^3$.

Let $v_i$, $i=1,\dots,k$ be the vectors generating the one-dimensional cones in $\Sigma(1)$, where $k=|\Sigma(1)|$. To each $v_i$ we associate an homogeneous coordinate $w_i \in \IC$. From the resulting $\IC^k$ we remove the set
\beq
Z_{\Sigma} = \bigcup_I \{(w_1,\dots,w_k):~w_i=0~\forall~i \in I\},
\eeq
where the union is taken over all sets $I \subseteq \{1,\dots,k\}$ for which $\{w_i:~i\in I\}$ does not belong to a cone in $\Sigma$. In other words, several $w_i$ are allowed to vanish simultaneously only if there is a cone such that the corresponding $v_i$ all belong to this cone.

Then the toric variety is given by 
\beq
\CM_{\Sigma}={\IC^k \setminus Z(\Sigma)\over G}
\eeql{toricvariety}
where $G$ is $(\IC^*)^{k-3}$ times a finite abelian group. For all the toric varieties we consider in these lectures the finite abelian group is trivial, so from now on we will omit it (see \cite{Skarke:1998yk} for an explanation of this group). The quotient by $(\IC^*)^{k-3}$ is implemented by taking equivalence classes with respect to the following equivalence relations among the coordinates $w_i$
\beq
(w_1,\ldots, w_k) \sim (\l^{Q_a^1}w_1,\ldots,\l^{Q_a^k}w_k)
\eeql{equiv1}
with $\lambda \in \IC^*$ and $\sum_{i=1}^k Q_a^i v_i =0$. Among these relations, $k-3$ are independent. We choose the $Q_a^i$ such that they are integer and the greatest common divisor of the $Q_a^i$ with fixed $a$ is $1$.

Using this construction, it is easy to see that the complex dimension of a toric variety is always equal to the real dimension $n$ of the lattice $N \cong \IZ^n$.

\begin{example}
Let us come back to the example of $\IC \IP^2$ (which is two-dimensional rather than three-dimensional, but easier to visualize as a first example). The fan is given by the first picture in figure \ref{CP2}. There are three one-dimensional cones generated by the vectors $v_1=(1,0)$, $v_2=(0,1)$ and $v_3=(-1,-1)$, to which we associate the homogeneous coordinates $w_1$, $w_2$ and $w_3$ of $\IC^3$. The set $Z_{\S}$ is simply $\{ 0 \}$, and thus the toric variety is given by
\beq
\cm_{\S} = (\IC^3 \setminus \{0\}) / (\IC^*).
\eeq
Moreover, we have that $1(1,0)+1(0,1)+1(-1,-1)=(0,0)$, so the $\IC^*$ quotient is implemented by the equivalence relation $(w_1,w_2,w_3) \sim \l (w_1,w_2,w_3)$. This is the usual description of $\IC \IP^2$.
\end{example}

\begin{example}
Now consider again $\IC \IP^{(2,3,1)}$. The fan is given by the second picture in figure \ref{CP2}. Again, $Z_{\S} = \{ 0 \}$, and 
\beq
\cm_{\S} = (\IC^3 \setminus \{0\}) / (\IC^*).
\eeq
But now, we have that $2(1,0)+3(0,1)+1(-2,-3)=(0,0)$, so the $\IC^*$ quotient is implemented by the equivalence relation $(w_1,w_2,w_3) \sim  (\l^2 w _1, \l^3 w_2,\l w_3)$.
\end{example}

\subsubsection{Properties}

We now state a few important properties of toric varieties, without proof. First, it is straightforward to know whether a toric variety is compact or not:

\begin{proposition}\label{compactness}
A toric variety $\cm_\S$ is compact if and only if its fan $\S$ fills $N_\IR$.
\end{proposition} 

The reader is referred to \cite{Fulton:1993} for a proof of this proposition, which will be very useful later on. For example, the fans of both $\IC \IP^2$ and $\IC \IP^{(2,3,1)}$, shown in figure \ref{CP2}, fill $N_\IR$; hence both manifolds are indeed compact.

It is also easy to see whether a toric variety is singular or not. We first need to define a few additional concepts. An $r$-dimensional cone is {\it simplicial} if it is generated by $r$ linearly independent one-dimensional vectors. We say that a fan is simplicial if all its cones are simplicial. Then, given a simplifical fan $\S$ it can be shown that the associated toric variety $\cm_\S$ can only have orbifold singularities. Moreover, if every $n$-dimensional cone of $\S$ is generated by vectors that generate the whole lattice $N$, then $\cm_\S$ is smooth. 

Hence, toric geometry provides a simple way to resolve orbifold singularities. Given a fan corresponding to a singular manifold, one can resolve the orbifold singularities by adding cones to the fan until every $n$-dimensional cone is generated by vectors generating $N$. This is one of the many aspects of toric geometry which is often used in string theory. However, we will not expand more on this in these notes; we simply give a quick example of how this works.

\begin{figure}[htp]
\begin{center}
\psfrag{v_1}{$v_1$}
\psfrag{v_2}{$v_2$}
\psfrag{v_3}{$v_3$}
\psfrag{v_4}{$v_4$}
\psfrag{v_5}{$v_5$}
\psfrag{v_6}{$v_6$}
\psfrag{WP2}{$\IC \IP^{(2,3,1)}$}
\psfrag{The resolved manifold}{The resolved manifold}
\includegraphics[width=12cm]{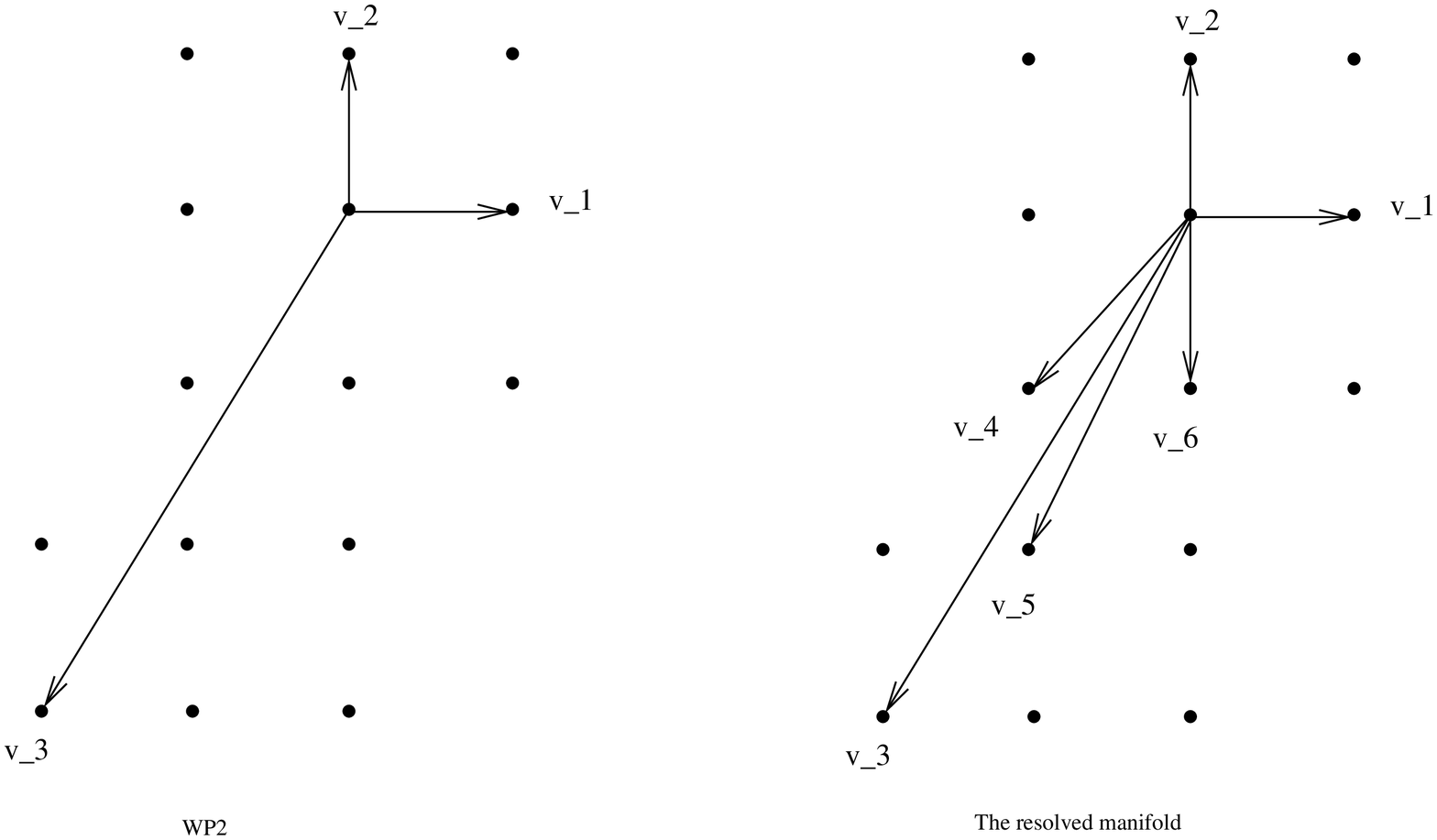}
\caption{The fan of $\IC \IP^{(2,3,1)}$ and its complete resolution.}
\label{f:sing}
\end{center}
\end{figure}

\begin{example}
Let us come back again to $\IC \IP^2$ and $\IC \IP^{(2,3,1)}$. First, in the case of $\IC \IP^2$, each two-dimensional cone in the fan is generated by vectors that generate $N$; hence $\IC \IP^2$ is smooth, as expected.

However the story is different for $\IC \IP^{(2,3,1)}$. The fan is given by the first picture of figure \ref{f:sing}. It is easy to see that the fan is simplicial (this is always the case for two-dimensional lattices), but that two of the three two-dimensional cones are not generated by vectors generating $N$. Hence $\IC \IP^{(2,3,1)}$ has orbifold singularities. In fact, one can check that it has two singularities, looking locally like $\IC^2 / \IZ_2$ and $\IC^2 / \IZ_3$ \cite{Skarke:1998yk}. To blow up these singularities, we add cones to the fan until all two-dimensional cones are generated by vectors generating $N$; we obtain the second picture of figure \ref{f:sing}, to which the associated manifold $\cm_\S$ is smooth. The $3$ extra toric divisors --- see the next subsection for a definition of toric divisors --- $v_4$, $v_5$ and $v_6$ that we added correspond to the blown up exceptional divisors.
\end{example}

\subsubsection{Toric divisors}

In a toric variety there is a natural set of divisors called toric divisors.

\begin{definition}
Let $\cm_\S$ be a toric variety described by a fan $\S$. As usual, associate an homogeneous coordinates $w_i$ to each vector $v_i$ generating the one-dimensional cones of $\S$. The {\it toric divisors} $D_i$ of $\cm_\S$ are the hypersurfaces defined by the equations $w_i=0$.
\end{definition}

Since we associated an homogeneous coordinates $w_i$ to each one-dimensional cones $v_i$ in the fan $\S$ of $\cm_\S$, we can think of the vectors $v_i$ as corresponding to the toric divisors defined by $w_i=0$.

Similarly, higher-dimensional cones of $\S$ correspond to lower dimensional algebraic subvarieties of $\cm_\S$.

In fact, it can be shown (see \cite{Fulton:1993}), by using methods very similar to those used for complex projective spaces, that the canonical bundle of $\cm_\S$ is given by
\beq
K_{\cm_\S} = \co(-\sum_i D_i).
\eeq
This result will be useful to determine whether a toric variety is \CY\ or not.

\subsection{Toric \CY\ threefolds}\label{CYcondition}

We will now implement the \CY\ condition on toric threefolds. 

We have seen in the first section that to a divisor $D=\sum_i a_i N_i$ we can associate a line bundle with a meromorphic section such that the meromorphic section has a zero of order $a_i$ along $N_i$ if $a_i > 0$ and a pole of order $-a_i$ along $N_i$ if $a_i < 0$. The $N_i$ are irreducible hypersurfaces, that is hypersurfaces that cannot be written as the union of two hypersurfaces.

In the toric case, the toric divisors $D_i$ defined by $w_i=0$ are irreducible hypersurfaces. Therefore, using the above correspondence we see that the toric divisor $D_i$ is associated to a line bundle $\co(D_i)$ with a section $s$ that has a zero of order one along $D_i$; thus the section $s$ is simply $w_i$. Hence we see that each homogeneous coordinate $w_i$ is a section of the line bundle $\co(D_i)$ associated to the toric divisor $D_i$.

Now, if we consider a monomial  $w_1^{a_1} \cdots w_k^{a_k}$; for $a_i>0$, it has zeroes of order $a_i$ along $D_i$, while for $a_j<0$, it has poles of order $-a_j$ along $D_j$. Therefore it is a section of the line bundle $\co(\sum_i a_i D_i)$.

Let us now consider the case where $a_i = \< v_i,m\>$, $i=1,\ldots,k$ for some $m \in M$. Under the equivalence relations of the toric variety the monomial becomes
\beq
(\l^{Q_a^1} w_1)^{\< v_1,m \>} \cdots (\l^{Q_a^k} w_k)^{\< v_k,m \>} = \l^{\< \sum_{i=1}^k Q_a^i  v_i, m\>} w_1^{\< v_1,m \>} \cdots w_k^{\< v_k,m \>}.
\eeq
But since $\sum_{i=1}^k Q_a^i  v_i=0$, this monomial is invariant under the equivalence relations and therefore it is a true, globally defined, meromorphic function on our toric variety. This means that it must be a section of the trivial line bundle, that is
\beq\label{equivdiv}
\sum_{i=1}^k \< v_i,m \> D_i \sim 0~~{\rm for~any}~m \in M.
\eeq
Conversely, if $\sum_{i=1}^k a_i D_i \sim 0$, then there exists a $m\in M$ such that $a_i= \< v_i,m \>$ for all $i$.

Now, we know that a K\"ahler manifold is \CY\ if and only if its canonical class is trivial. We saw in the previous section that the canonical line bundle of a toric variety $\cm_{\S}$ is given by $K_{\CM_{\Sigma}} \cong \CO (-\sum_{i=1}^k D_i)$. Therefore the canonical bundle is trivial if and only if $\sum_{i=1}^k D_i \sim 0$. Using \refeq{equivdiv}, we see that this condition is equivalent to the existence of a $m \in M$ such that $\< v_i,m \>=1$ for all $i$, which leads to the following proposition.

\begin{proposition}\label{affine}
Let $\cm_\S$ be a toric manifold defined by a fan $\S$. $\cm_\S$ is \CY\ if and only if the vectors $v_i$ generating the one-dimensional cones of $\cm_\S$ all lie in the same affine hyperplane.
\end{proposition}

It is thus very easy to see whether a toric variety is \CY\ or not; in fact, it can be read off directly from the fan $\S$ of the toric variety. For instance, according to this proposition it is clear that $\IC \IP^2$ and $\IC \IP^{(2,3,1)}$ are not \CY, as expected.

A consequence of proposition \ref{affine} is the following:

\begin{corollary}
A toric \CY\ manifold is noncompact.
\end{corollary}

Since the $v_i$ lie in a hyperplane, $\Sigma$ does not fill $N_{\IR}$. Thus proposition \ref{compactness} tells us that $\cm_\S$ is noncompact.

This seems like a serious limitation of toric geometry, since in string theory we are often interested in compact \CY\ manifolds. However, we will see in section \ref{hyper} how to construct compact \CY\ manifolds in toric geometry.

The \CY\ condition can be rewritten in yet another equivalent form. In \refeq{equiv1} we defined the `charges' (the meaning of this name will become clear in section \ref{sympl}) $Q_a^i$ satisfying $\sum_{i=1}^k Q_a^i v_i =0$. Therefore $\sum_{i=1}^k Q_a^i \< v_i,m \> =0$ for any $m \in M$. In particular, there exists an $m \in M$ such that $\< v_i,m \>=1$ for all $i$ if and only if $\sum_{i=1}^k Q_a^i = 0$ for all $a$. But we showed that a toric manifold is \CY\ if and only if there exists and $m \in M$ such that $\< v_i,m \>=1$ for all $i$. Therefore, the condition can be restated as follows: 

\begin{proposition}
A toric manifold is \CY\ if and only if the charges $Q_a^i$ satisfy the condition $\sum_{i=1}^k Q_a^i = 0$ for all $a$.
\end{proposition}

This condition is also very simple to verify. We only have to check that the charges $Q^i_a$ given in the toric data describing the manifold add up to zero. Thus, if we are given a fan we simply check that the $v_i$ lie in an affine hyperplane, while if we are given the toric data we simply verify that the charges add up to zero.

To conclude this section we introduce a nice pictorial way of characterizing toric \CY\ threefolds. We showed that for toric \CY\ threefolds the $v_i$ lie in a two-dimensional plane $P$. Therefore, we can draw the two-dimensional graph ${\tilde \Gamma}$ given by the intersection of the plane $P$ and the fan $\Sigma$. ${\tilde \Gamma}$ determines completely the fan $\Sigma$ of a toric \CY\ threefold. Given ${\tilde \Gamma}$, we can draw a `dual' graph $\Gamma$ in the sense that the edges of ${\tilde \Gamma}$ are normals to the edges of $\Gamma$ and vice-versa. $\Gamma$ is called the {\it toric diagram} of a toric \CY\ threefolds $\CM_{\Sigma}$. It represents the degeneration of the fibers of the torus fibration. We will describe in more details toric diagrams in section \ref{sympl}.

\begin{figure}[htp]
\begin{center}
\psfrag{(-1,-1)}{$(-1,-1)$}
\psfrag{(0,0)}{$(0,0)$}
\psfrag{(1,0)}{$(1,0)$}
\psfrag{(0,-1)}{$(0,-1)$}
\includegraphics[width=6cm]{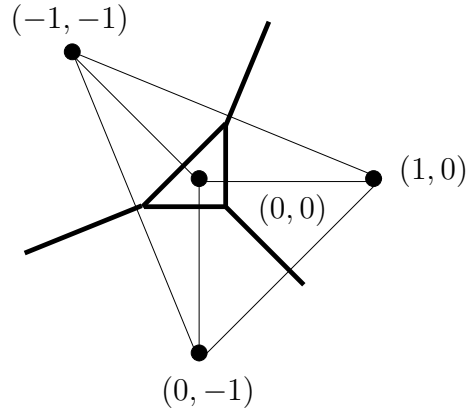}
\caption{The $\Gamma$ and $\tilde \Gamma$ graphs for $\CO(-3) \rightarrow \IC \IP^2$. The toric diagram $\Gamma$ is the normal diagram drawn in thick lines. The points $(v_i,1)$ give the fan $\Sigma$, where the $v_i$ are the vertices of $\tilde \Gamma$ and are shown in the figure.}
\label{p2graphs}
\end{center}
\end{figure}

Conversely, given a toric diagram $\Gamma$, it is straightforward to recover the fan $\Sigma$ of the toric \CY\ threefold. One first draws the dual graph $\tilde \Gamma$, and then define the vectors $v_i = (\nu_i,1)$ where $\nu_i$ are the vertices of $\tilde \Gamma$. Because of the symmetries of a three-dimensional lattice, the $v_i$ must be the generators of the edges of the fan $\Sigma$ of the toric \CY\ threefold $\CM_{\Sigma}$. Linear relations between the vectors $v_i$ give the charges $Q_a^i$. In other words, the fan $\Sigma$ is a three-dimensional cone over the two-dimensional graph $\tilde \Gamma$. An example of graphs $\Gamma$ and $\tilde \Gamma$ is given in figure \ref{p2graphs}.

\subsection{Toric diagrams and symplectic quotients}\label{sympl}

In this section we describe the toric diagrams introduced above. To do so, we need to leave momentarily the homogeneous coordinates approach to toric varieties and see toric manifolds as symplectic quotients, or correspondingly as the Higgs branch of the space of supersymmetric vacua of the gauged linear sigma model.

\subsubsection{Toric manifolds as symplectic quotients}

Let $z_1,\dots,z_k$ be the coordinates of $\IC^k$. Let $\mu_a: \IC^k \to \IC$, $a=1,\dots,k-3$ be the $k-3$ moment maps defined by
\beq
\sum_{i=1}^k Q_a^i |z_i|^2 = \Re(t_a),
\eeql{moment}
where the $t_a$ are complex numbers. The $Q_a^i$ are the same charges that were introduced in \refeq{equiv1}. Therefore, the \CY\ condition imposes that $\sum_{i=1}^k Q_a^i=0$ for all $a$. We also consider the action of the group $G=U(1)^{k-3}$ on the coordinates defined by
\beq
z_j \rightarrow \exp (i Q_a^j \alpha_a ) z_j,~~~a=1,\dots,k-3.
\eeq
It turns out that
\beq
\CM = {\bigcap_{a=1}^{k-3} \mu^{-1} (\Re(t_a)) \over G}
\eeq
is a toric \CY\ threefold. The $k-3$ parameters $t_a$ are the complexified K\"ahler parameters of the \CY\ threefold.

Furthermore, since the charges $Q_a^i$ are the same as in \refeq{equiv1}, it is easy to recover the fan of $\CM$. One only has to find distinct vectors $v_i$ satisfying $\sum_{i=1}^k Q_a^i v_i =0 $; the $v_i$ generate the one-dimensional cones of $\Sigma$. Moreover, since the \CY\ condition tells us that $\sum_{i=1}^k Q_a^i  =0$, we can choose (because of the symmetries of three-dimensional lattices) vectors $v_i$ of the form $v_i = (\nu_i,1)$. The problem is then reduced to a two-dimensional problem which can easily be solved by inspection. We see that the charges $Q_a^i$ are the important data defining the \CY\ toric manifolds. This is usually called the {\it toric data} of the manifold.

This description of toric manifolds also arise in gauged linear sigma models. This is a two-dimensional $U(1)^{k-3}$ gauge theory with $k$ chiral superfields $\Phi_i$, whose scalar components are the $z_k$. The charges of the superfields $\Phi_i$ under the gauge group $U(1)^{k-3}$ are denoted by $Q_a^i$, $a=1,\dots,k-3$. This is why the $Q_a^i$ are generally called {\it charges}. It turns out that --- in the Higgs branch --- the supersymmetric ground states of the theory are parameterized by the so-called D-term equations modulo gauge equivalence, which are nothing but the moment maps $\mu_a$ defined in \refeq{moment}. In other words, the Higgs branch of the space of supersymmetric ground states of the gauged linear sigma model is the toric variety $\CM$ defined above.

Now equipped with the description of toric \CY\ threefolds as symplectic quotients, let us come back to the toric diagrams introduced in section \ref{CYcondition}. There, we claimed that these diagrams encode the degenaration of the fibers of the manifold. This can be seen in two different ways: by looking at the threefold as a $T^3$ fibration or as a $T^2 \times \IR$ fibration. We will start with the first approach in section \ref{t3}, which is probably simpler. We will explore the second point of view in section \ref{t2}, using the topological vertex approach to toric \CY\ threefolds.

\subsubsection{$T^3$ fibration}\label{t3}

We look at the threefolds as $T^3$ fibrations over three dimensional base manifolds with corners. Locally, we can introduce complex coordinates on the toric manifold: these are the $z_i$ introduced in \refeq{moment}. They are not all independent; for a threefold, there are $k-3$ relations between them given by the moment maps \refeq{moment}. Let us rewrite these coordinates as $z_j=|z_j| e^{i \theta_j}$, and introduce a new set of coordinates $\{ (p_1,\theta_1),\dots,(p_k,\theta_k)\}$, with $p_i\equiv |z_i|^2$, $i=1,\ldots,k$. The base of the threefold is then parameterized by the coordinates $p_i$, while the phases $\theta_i$ describe the fiber $T^3$. 

Since $|z_i|^2 \geq 0$, the coordinates $p_i$ satisfy $p_i \geq 0$. Therefore the boundaries of the base are where some of the coordinates $p_i$ vanish. But when $p_j=0$ the circle $|z_j|e^{i \theta_j}$ degenerates to a single point. Hence, the boundaries of the base correspond to degenerations of the corresponding fiber directions $\theta_j$. Geometrically, this means that the fiber degenerates in the direction given by the unit normal to the boundary.

To draw the toric diagram, we first use the moment maps \refeq{moment} to express the coordinates $p_j$, $j=4,\ldots,k$ in terms of the three coordinates $p_1$, $p_2$, $p_3$. Consequently, the boundary equations $p_j=0$, $j=4,\ldots,k$ become equations in the coordinates $p_1$, $p_2$ and $p_3$ involving the K\"ahler parameters $t_j$ of \refeq{moment}. In fact, each boundary equation gives a plane in the space generated by $p_1$, $p_2$ and $p_3$. The intersections of these planes are lines; they form the toric diagram of the toric variety, visualized as a three dimensional graph in the space generated by $p_1$, $p_2$ and $p_3$.

Hence, in this approach the toric diagram is simply the boundary of the three dimensional base parameterized by the $p_i$. There is a $T^3$ fiber over the generic point, which degenerates at the boundaries in a way determined by the unit normal. Thus, from this point of view toric diagrams should be visualized as three dimensional diagrams, encoding the degeneration of the $T^3$ fiber. It is perhaps simpler to understand this approach by working out a specific example.

\begin{figure}[htp]
\begin{center}
\psfrag{z1}{$p_1$}
\psfrag{z2}{$p_2$}
\psfrag{z3}{$p_3$}
\includegraphics[width=6cm]{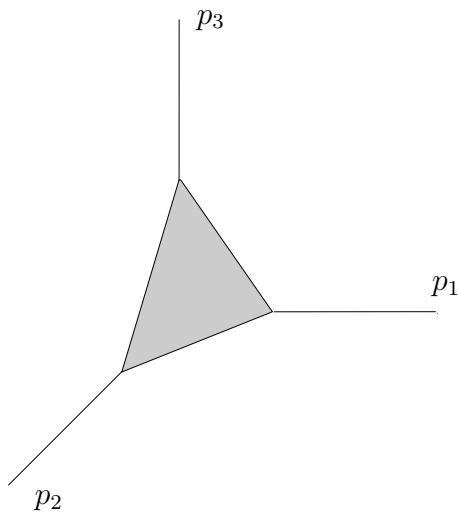}
\caption{Toric diagram $\Gamma$ of $\CO(-3) \rightarrow \IC \IP^2$ visualized as a three dimensional graph. It encodes the degeneration loci of the $T^3$ fiber.}
\label{p2graph3d}
\end{center}
\end{figure}

\begin{example}\label{exampcp2}
Let us find the toric diagram of $\CO(-3) \rightarrow \IC \IP^2$ from this point of view. This manifold is defined by the moment map $p_1+p_2+p_3-3p_4=t$, which we can use to express $p_4={1\over 3} (p_1+p_2+p_3-t)$. The boundary planes are then given by $p_1=0$, $p_2=0$, $p_3=0$ and $p_1+p_2+p_3=t$. The intersections of these planes give the toric diagram of $\CO(-3) \rightarrow \IC \IP^2$, which is drawn in figure \ref{p2graph3d}. We see that it is the same toric diagram as the one shown in figure \ref{p2graphs}, but visualized as a three dimensional graph. Note that from the fourth boundary equation one can see that the K\"ahler parameter $t$ controls the size of the $\IC \IP^2$, as it should be.
\end{example}

This is indeed an easy way to visualize the geometry of the manifold from the toric diagram; another example of this approach will be given in section \ref{examples}. However, it turns out that in many situations it is more enlightening to consider the manifold as a $T^2 \times \IR$ fibration, especially from the topological vertex perspective. Let us now describe this alternative viewpoint.

\subsubsection{$T^2 \times \IR$ fibration}\label{t2}

In this language, a toric diagram $\Gamma$ is a two-dimensional graph which represents the degeneration locus of the $T^2 \times \IR$ fibration over the base $\IR^3$. Over a line in $\Gamma$ in the direction $(q,p)$, the cycle $(-q,p)$ of the $T^2$ fiber degenerates. 

To exhibit this structure, we will now follow the topological vertex approach to toric \CY\ threefolds developed by Aganagic, Klemm, Mari\~no and Vafa in \cite{Aganagic:2003db}. A good review is \cite{Marino:2004uf}.

The fundamental idea behind this approach is that toric \CY\ threefolds are built by gluing together $\IC^3$ patches. Therefore, the first step is to describe $\IC^3$ (which is the simplest noncompact toric \CY\ threefold) as a $T^2 \times \IR$ fibration and exhibit its degeneration locus in a two-dimensional graph $\Gamma$, which turns out to be a trivalent vertex. Then, more general geometries are constructed by gluing together $\IC^3$ patches, which, in the toric diagram language, corresponds to gluing together trivalent vertices in a way specified by the toric data of the manifold. 

Conversely, given a toric \CY\ threefold, we can find a decomposition of the set of all coordinates into triplets that correspond to the decomposition of the threefold into $\IC^3$ patches. The moment maps \refeq{moment} relate the coordinates between the patches, therefore describing how the trivalent vertices corresponding to the $\IC^3$ patches are glued together to form the toric diagram of the manifold.

Let us start by describing $\IC^3$ from this point of view. Here we will only sketch the description; the details are given in \cite{Marino:2004uf,Aganagic:2003db}. Let $z_i$, $i=1,2,3$ be complex coordinates on $\IC^3$. Define the functions 
\bea
&&r_{\alpha}(z)=|z_1|^2-|z_3|^2, \nn \\ 
&&r_{\beta}(z)=|z_2|^2-|z_3|^2, \nn \\
&&r_{\g} (z)=\Im(z_1 z_2 z_3).
\eeal{hamil}
It turns out that these functions generate the fiber $T^2 \times \IR$. More specifically, $\IR$ is generated by $r_{\g}$ while the $T^2$ fiber is generated by the circle actions
\beq
\exp (i\a r_{\a} + i \b r_{\b}): (z_1,z_2,z_3) \rightarrow (e^{i \a} z_1, e^{i \b} z_2, e^{-i(\a+\b)} z_3).
\eeql{cycles}
The cycles generated by $r_{\a}$ and $r_{\b}$ are then respectively referred to as the $(0,1)$ and $(1,0)$ cycles.

We now describe the degeration loci of the fibers. We see from \refeq{hamil} and \refeq{cycles} that the $(0,1)$ cycle degenerates when $r_{\a} =0=r_{\g} $ and $r_{\b} \geq 0$, while the $(1,0)$ cycle degenerates when $r_{\a} \geq 0=r_{\g}$ and $r_{\b} = 0$. There is also a one-cycle parameterized by $\a+\b$ that degenerates when $r_{\a}-r_{\b} =0=r_{\g}$ and $r_{\a} \leq 0$.

\begin{figure}[htp]
\begin{center}
\psfrag{(0,1)}{$(0,1)$}
\psfrag{(-1,-1)}{$(-1,-1)$}
\psfrag{(1,0)}{$(1,0)$}
\includegraphics[width=5cm]{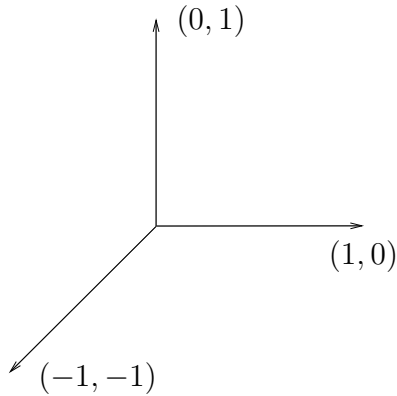}
\caption{Trivalent vertex associated to $\IC^3$, drawn in the $r_{\alpha}$-$r_{\b}$ plan.  The vectors represent the generating cycles over the lines.}
\label{trivertex}
\end{center}
\end{figure}

The toric diagram is a planar graph that encodes the degeneration loci of the fibers. We can set $r_{\g}=0$ and draw the graph in the plane $r_{\a}-r_{\b}$. The graph consists in lines $p r_{\a}+q  r_{\b} = c$ where $c$ is a constant. Over this line the $(-q,p)$ cycle of the $T^2$ fiber degenerates (up to the equivalence $(q,p) \sim (-q,-p)$). For $\IC^3$, the degeneration loci can be represented as a toric diagram with lines defined by the equations $r_{\a}=0,~r_{\b}\geq 0$; $r_{\b}=0,~r_{\a} \geq 0$ and $r_{\a}-r_{\b} =0,~r_{\a} \leq 0$. Over these lines respectively the cycles $(0,1)$; $(-1,0) \sim (1,0)$ and $(1,1)$ degenerate. This gives the trivalent vertex associated to $\IC^3$, which is shown in figure \ref{trivertex}.

For more general geometries, we first find a decomposition of the set of coordinates $z_i$, $i=1,\dots,k$ into triplets of coordinates associated to the $\IC^3$ patches. We choose a patch and describe the functions $r_{\a}$ and $r_{\b}$ as above. It turns out that we can use these coordinates as global coordinates for the $T^2$ fiber in the $\IR^3$ base. As usual, we refer to the cycles $r_{\a}$ and $r_{\b}$ respectively as the $(0,1)$ and $(1,0)$ cycles. Using the moment maps \refeq{moment} defining the toric \CY\ threefold, we can find the action of the functions $r_{\a}$ and $r_{\b}$ on the other patches and therefore draw the toric diagram giving the degeneration loci of the $T^2$ fiber. An explicit example of this approach will be worked out in section \ref{examples}.

This decomposition of toric \CY\ threefolds into $\IC^3$ patches leads to a similar decomposition of topological string amplitudes on toric \CY\ threefolds into a basic building block associated to the trivalent vertex of the $\IC^3$ patches, which is called the {\it topological vertex}. By gluing together these topological vertices one can build topological string amplitudes on any toric \CY\ threefold. This beautiful property of topological amplitudes is the essence of the topological vertex approach developed in \cite{Aganagic:2003db}. We will not expand further on this subject; the interested reader is encouraged to go through the details of the construction in \cite{Aganagic:2003db}.

In the next section we illustrate these different approaches to toric \CY\ threefolds in specific examples.

\subsection{Examples}\label{examples}

We now describe two examples of toric \CY\ threefolds. The first example is the resolved conifold, namely $\CO(-1) \oplus \CO(-1) \rightarrow \IC \IP^1$. In this simple case, we illustrate in details the different viewpoints explained in the previous sections. The second example is a more complicated geometry. It is a noncompact \CY\ threefold whose compact locus consists of two compact divisors each isomorphic to a del Pezzo surface $dP_2$\footnote{A del Pezzo surface $dP_n$, $n=0,\ldots,8$ is a complex two-dimensional Fano variety, which can be understood as $\IC \IP^2$ blown up in $n$ points in general position.} and a rational $(-1,-1)$ curve that intersects both divisors transversely. We will give the toric data describing the manifold and draw the corresponding toric diagram.

\subsubsection{$\CO(-1) \oplus \CO(-1) \rightarrow \IC \IP^1$}

The resolved conifold $Y={\cal O}(-1) \oplus {\cal O}(-1) \rightarrow \IC {\IP}^1$  is a noncompact \CY\ threefold which admits a toric description given by the following toric data:
\beq
\begin{matrix}
 &  z_1 & z_2 & z_3 & z_4  \\
\IC^* & 1 & 1 & -1 & -1
\end{matrix}
\eeql{toricres}
The lines in this table give the charges $Q^i_a$ corresponding to the torus actions on the homogeneous coordinates $z_i$. We see that $\sum_i Q^i = 1+1-1-1=0$; therefore $Y$ is \CY. $Y$ is defined as the space obtained from
\beq
|z_1|^2 + |z_2|^2 -|z_3|^2 - |z_4|^2 =t
\eeql{rescon}
after quotienting by the $U(1)$ action specified by the charges in \refeq{toricres}.

We now find the fan $\Sigma$ describing $Y$. We have the relation $\sum_{i=1}^4 Q^i v_i = v_1+v_2-v_3-v_4=0$. We choose distinct vectors $v_i = (w_i,1)$ where $w_i$ is two-dimensional. A solution is $v_1 = (1,0,1)$, $v_2=(-1,0,1)$, $v_3=(0,1,1)$ and $v_4=(0,-1,1)$. These four vectors generate the four one-dimensional cones of $\Sigma$.

\begin{figure}[htp]
\begin{center}
\psfrag{(0,1)}{$(0,1)$}
\psfrag{(0,-1)}{$(0,-1)$}
\psfrag{(-1,0)}{$(-1,0)$}
\psfrag{(1,0)}{$(1,0)$}
\includegraphics[width=7cm]{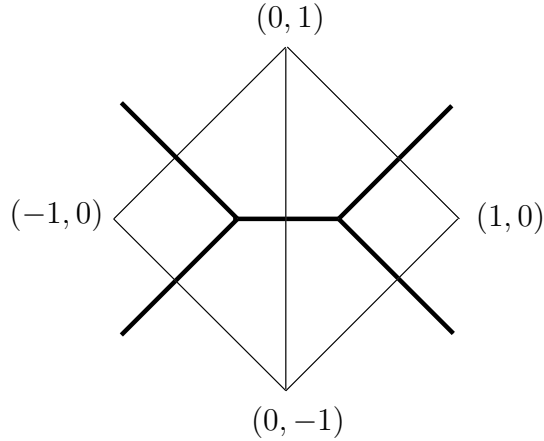}
\caption{The $\Gamma$ and $\tilde \Gamma$ graphs for $\CO(-1)\oplus \CO(-1) \rightarrow \IC \IP^1$. The toric diagram $\Gamma$ is the normal diagram drawn in thick lines. The points $(v_i,1)$ give the fan $\Sigma$, where the $v_i$ are the vertices of $\tilde \Gamma$ and are shown in the figure.}
\label{p1graphs}
\end{center}
\end{figure}

The two-dimensional graph $\tilde \Gamma$ is given by the intersection of the plane $z=1$ and $\Sigma$. The vertices are $(1,0)$,$(-1,0)$,$(0,1)$ and $(0,-1)$. We can also draw the toric diagram, which is the dual graph $\Gamma$. They are shown in figure \ref{p1graphs}.

If we look at the resolved conifold as a $T^3$ fibration, we have to understand the toric diagram $\Gamma$ as a three-dimensional graph representing the base, where the $T^3$ fiber degenerates at the boundaries. The base is parameterized by the four coordinates $p_i \equiv |z_i|^2$ subject to the relation \refeq{rescon}. We can use \refeq{rescon} to eliminate $p_4$,
\beq
p_4=p_1+p_2-p_3-t.
\eeq 
Therefore, since $|z_i|^2 \geq 0$, the boundary equations of the toric base are given by
\bea
p_1 &=& 0, \nn\\
p_2 &=& 0, \nn\\
p_3 &=& 0, \nn\\
p_1+p_2-p_3 &=& t.
\eea

\begin{figure}[htp]
\begin{center}
\psfrag{(0,0,t)}{$(0,0,t)$}
\psfrag{p1}{$p_1$}
\psfrag{p2}{$p_2$}
\psfrag{p3}{$p_3$}
\includegraphics[width=7cm]{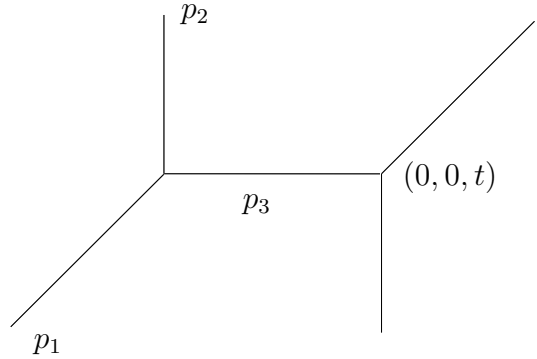}
\caption{Toric diagram $\Gamma$ of $\CO(-1)\oplus \CO(-1) \rightarrow \IC \IP^1$ visualized as a three-dimensional graph. It encodes the degeneration loci of the $T^3$ fiber.}
\label{p1graph3d}
\end{center}
\end{figure}

The intersections of these planes give the toric diagram of the resolved conifold shown in figure \ref{p1graphs}, but visualized as a three dimensional graph as in figure \ref{p1graph3d}. Note that as in example \ref{exampcp2}, by the fourth boundary equation above one can see that the K\"ahler parameter $t$ controls the size of the $\IC \IP^1$, as it should be.

We can also describe the resolved conifold as a $T^2 \times \IR$ fibration, using its decomposition into $\IC^3$ patches. We choose the first patch to be defined by $z_1 \neq 0$. Using \refeq{rescon} we can express $z_1$ in terms of the other coordinates, so the patch is parameterized by $(z_2,z_3,z_4)$. We define the functions
\bea
r_{\a}&=&|z_3|^2-|z_2|^2,\nn\\
r_{\b}&=&|z_4|^2-|z_2|^2.
\eeal{hamilp11}
This gives the usual trivalent graph of $\IC^3$.

\begin{figure}[htp]
\begin{center}
\psfrag{(0,1)}{$(0,1)$}
\psfrag{(1,0)}{$(1,0)$}
\psfrag{(-1,-1)}{$(-1,-1)$}
\psfrag{(1,1)}{$(1,1)$}
\psfrag{(0,-1)}{$(0,-1)$}
\psfrag{(-1,0)}{$(-1,0)$}
\psfrag{U1}{$U_1$}
\psfrag{U2}{$U_2$}
\includegraphics[width=8cm]{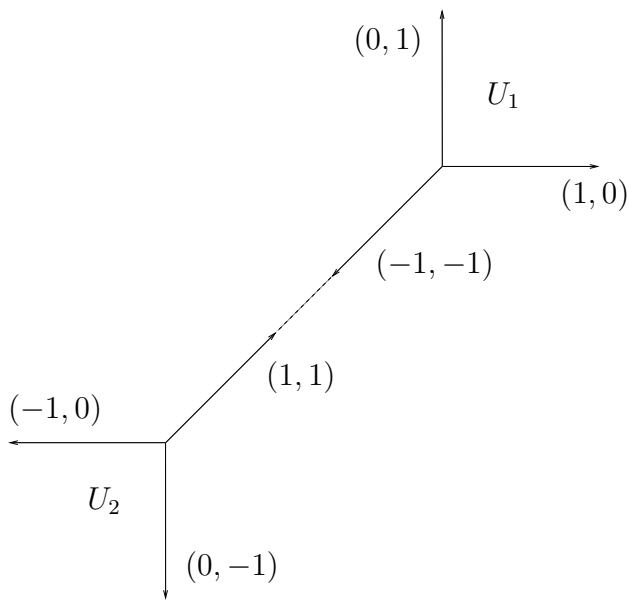}
\caption{Toric diagram of $\CO(-1)\oplus\CO(-1)\rightarrow\IC\IP^1$, drawn in the $r_{\alpha}$-$r_{\b}$ plan.  The vectors represent the generating cycles over the lines. The origin of the second patch $U_2$ is shifted to $(-t,-t)$.}
\label{p1graphtopo}
\end{center}
\end{figure}

The other patch is defined by $z_2 \neq 0$, therefore parameterized by $(z_1,z_3,z_4)$. Using \refeq{rescon} we can rewrite the functions \refeq{hamilp11} in terms of the coordinates on this patch:
\bea
r_{\a}&=&|z_1|^2-|z_4|^2+t,\nn\\
r_{\b}&=&|z_1|^2-|z_3|^2+t.
\eeal{hamilp12}
These functions generate the circle action
\beq
\exp (i\a r_{\a} + i \b r_{\b}): (z_1,z_3,z_4) \rightarrow (e^{i (\a+\b)} z_1, e^{-i \b} z_3, e^{-i\a} z_4).
\eeq
In this patch, the $(0,1)$ cycle degenerates when $r_{\a} \leq -t$ and $r_{\b} = -t$. The $(1,0)$ cycle degenerates when $r_{\a}=-t$ and $r_{\b} \leq -t$. The $(1,1)$ cycle degenerates when $r_{\a}-r_{\b}=0$ and $r_{\a} \geq -t$. Therefore, the graph associated to this patch is identical to the first one, although it is shifted such that its origin is at the point $(-t,-t)$. The two graphs are joined through the common edge given by $r_{\a}-r_{\b}=0$. $t$ gives the `length' of the internal edge, and correspondingly is the K\"ahler parameter associated to the $\IC \IP^1$. This gives the toric diagram of the resolved conifold shown in figure \ref{p1graphtopo}.

\subsubsection{Two $dP_2$'s connected by a $\IC \IP^1$}

We now present a more complicated example, which is a noncompact \CY\ threefold $X$ whose compact locus consists of two compact divisors each isomorphic to a del Pezzo surface $dP_2$ and a rational $(-1,-1)$ curve that intersects both divisors transversely. The divisors do not intersect each other. This manifold is described by the following toric data:
\beq
\begin{matrix} 
 & z_1 & z_2 & z_3 & z_4 & z_5 & z_6 & z_7 & z_8 & z_9 & z_{10} \\
\IC^* & -1 & 1 & 1 & -1 & 0 & 0 & 0 & 0 & 0 & 0 \\
\IC^* & 1 & 0 & -1 & -1 & 0 & 1 & 0 & 0 & 0 & 0 \\
\IC^* & 1 & -1 & 0 & -1 & 1 & 0 & 0 & 0 & 0 & 0 \\
\IC^* & 0 & 0 & 0 & 1 & -1 & -1 & 1 & 0 & 0 & 0 \\
\IC^* & 0 & 0 & 0 & 0 & 1 & 0 & -1 & -1 & 0 & 1 \\
\IC^* & 0 & 0 & 0 & 0 & 0 & 1 & -1 & 0 & -1 & 1 \\
\IC^* & 0 & 0 & 0 & 0 & 0 & 0 & -1 & 1 & 1 & -1.
\end{matrix}
\eeql{toricXi}

\begin{figure}[htp]
\begin{center}
\includegraphics[width=7cm]{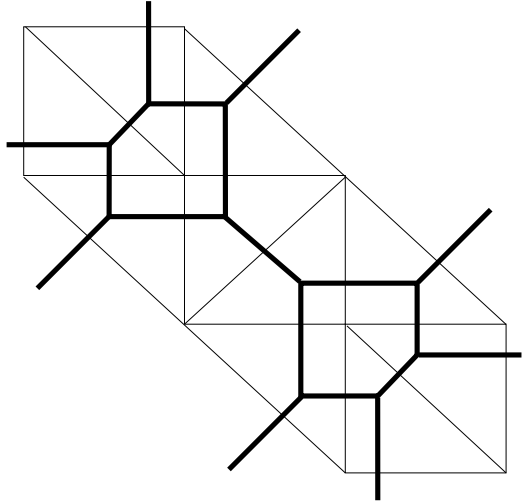}
\caption{The $\Gamma$ and $\tilde \Gamma$ graphs for the \CY\ threefold $X$ whose compact locus consists of two $dP_2$'s connected by a $\IC \IP^1$. The toric diagram $\Gamma$ is the normal diagram drawn in thick lines.}
\label{maingeograph}
\end{center}
\end{figure}

We see that the charges in each line add up to zero, hence $X$ is \CY. The toric diagram $\Gamma$ of $X$ and its dual $\tilde \Gamma$ are shown in figure \ref{maingeograph}. 

\subsection{Hypersurfaces in toric varieties}\label{hyper}

In the remaining of this section we explain how compact \CY\ manifolds may be obtained in toric geometry, namely as hypersurfaces in compact toric manifolds using Batyrev's well known reflexive polytopes \cite{Batyrev:1994}.

\subsubsection{Reflexive polytopes}
 
In section \ref{toricCY} we described in details toric \CY\ threefolds. In particular, we showed that toric \CY\ threefolds are noncompact. However, from a string theory perspective, it is often desirable to consider compact \CY\ manifolds. Hence it seems that toric geometry is not a good setup for such geometries. 

Fortunately, there is a way to construct compact \CY\ manifolds in toric geometry, namely as compact hypersurfaces in compact toric varieties. Batyrev's reflexive polytopes \cite{Batyrev:1994} provide a very useful description of such compact \CY\ manifolds. The toric variety itself is not \CY\, consequently it can be compact. Reflexivity of the polytopes then ensures that the compact hypersurface, which is not toric itself, is \CY.

An elementary introduction to these concepts and their applications to string theory and dualities can be found in \cite{Skarke:1998yk}. The following is partly based on the first sections of \cite{Bouchard:2003bu}.

As in section \ref{toricCY}, in the following we focus on three-dimensional toric varieties, therefore leading to two-dimensional \CY\ hypersufaces, i.e. K3 surfaces. It is however straightforward to generalize the concepts to higher dimensional toric varieties.

A polytope in $M_\IR$ is the convex hull of a finite number of points in $M_\IR$, and a polyhedron in $M_\IR$ is the intersection of finitely many half-spaces (given by inequalities $\langle u,v\rangle \geq c$ with some $v\in N_\IR$ and $c\in \IR$) in $M_\IR$. It is well known that any polytope is a polyhedron and any bounded polyhedron is a polytope. If a polyhedron $\D \subset M_\IR$ contains the origin \ipo, its dual
\beq
\D^*=\{v\in N_\IR : \langle u,v\rangle \geq -1 \hbox{ for all } u \in \D \}.
\eeql{dualpoly}
is also a polyhedron containing \ipo, and $(\D ^*)^*=\D $.

A {\it lattice polytope} in $M_\IR$ is a polytope with vertices in $M$. 

\begin{definition}
A polytope $\D\subset M_\IR$ containing \ipo\ is called {\rm reflexive} if both $\D$ and $\D^*$ are lattice polytopes. 
\end{definition}

This is equivalent to $\D$ being a lattice polytope whose bounding equations are of the form $\<u,v_i\>\ge -1$ with $v_i\in N$ (in coordinates, $\sum_j u_j v_{ij}\ge -1$ with integer coefficients $v_{ij}$). By convexity it is sufficient to consider only those equations corresponding to $v_i$ that are vertices of $\D^*$. In this way there is a duality between vertices of $\D^*$ and facets of $\D$; similarly, there are dualities between $p$-dimensional faces of $\D$ and $(n-p-1)$-dimensional faces of $\D^*$ (in three dimensions: between edges and dual edges).

An interior point $u$ of a reflexive polytope must satisfy $\langle u,v_i\rangle > -1$ for all $v_i$, so an interior lattice point must satisfy $\langle u,v_i\rangle \geq 0$. Thus if $u$ is an interior lattice point, then $nu$ is also an interior lattice point for any non-negative integer $n$. For $u\ne$ \ipo\ this would be in conflict with the boundedness of $\D$, implying that \ipo\ is the only interior lattice point.

\subsubsection{Toric interpretation}\label{Kthree}

Given a pair of three dimensional reflexive polytopes $\D\in M_\IR$, $\D^*\in N_\IR$, a smooth K3 surface can be constructed in the following way. Any complete triangulation of the surface of $\D^*$ defines a fan $\Sigma$ whose three dimensional cones are just the cones over the regular (i.e., lattice volume one) triangles. To any lattice point $p_i=(\bar x_i, \bar y_i, \bar z_i)$ on the boundary of $\D^*$ one can assign a homogeneous coordinate $w_i\in \IC$, with the rule that several $w_i$ are allowed to vanish simultaneously only if there is a cone such that the corresponding $p_i$ all belong to this cone. The equivalence relations among the homogeneous coordinates are given by
\beq
(w_1,\ldots, w_n) \sim (\l^{Q_a^1}w_1,\ldots,\l^{Q_a^k}w_k)\;\hbox{ for any }\l\in\IC^*
\eeql{equiv}
with any set of integers $Q_a^i$ such that $\sum Q_a^i p_i=0$; among these relations, $k-3$ are independent. This construction gives rise to a smooth compact three dimensional toric variety $\CM_{\Sigma}$ (smooth because the generators of every cone are also generators of $N$, compact because the fan fills $N_\IR$). The loci $w_i=0$ are the toric divisors $D_i$.

To any lattice point $q_j$ of $M$ we can assign a monomial $m_j=\prod_i w_i^{\<q_j,p_i\>+1}$; the exponents are non-negative as a consequence of reflexivity. The hypersurface defined by the zero-locus of a generic polynomial $P=\sum a_j m_j$ transforms homogeneously under \refeq{equiv} and can be shown to define a K3 hypersurface in $\CM_{\Sigma}$ (actually it defines a family of hypersurfaces depending on the coefficients $a_j$).

\begin{remark}
A good way to remember this construction is to note that the polytope in $\D^* \in N_\IR$ gives the faN of the ambient toric variety, while the polytope in $\D \in M_\IR$ gives the MonoMials.
\end{remark}

\begin{figure}[htp]
\begin{center}
\psfrag{D}{$\D$}
\psfrag{Dstar}{$\D^*$}
\psfrag{x3}{$x^3$}
\psfrag{y3}{$y^3$}
\psfrag{z3}{$z^3$}
\psfrag{xyz}{$xyz$}
\psfrag{x2y}{$x^2 y$}
\psfrag{xy2}{$x y^2$}
\psfrag{y2z}{$y^2 z$}
\psfrag{yz2}{$y z^2$}
\psfrag{z2x}{$x^2 z$}
\psfrag{zx2}{$x z^2$}
\psfrag{x}{$x$}
\psfrag{y}{$y$}
\psfrag{z}{$z$}
\includegraphics[width=12cm]{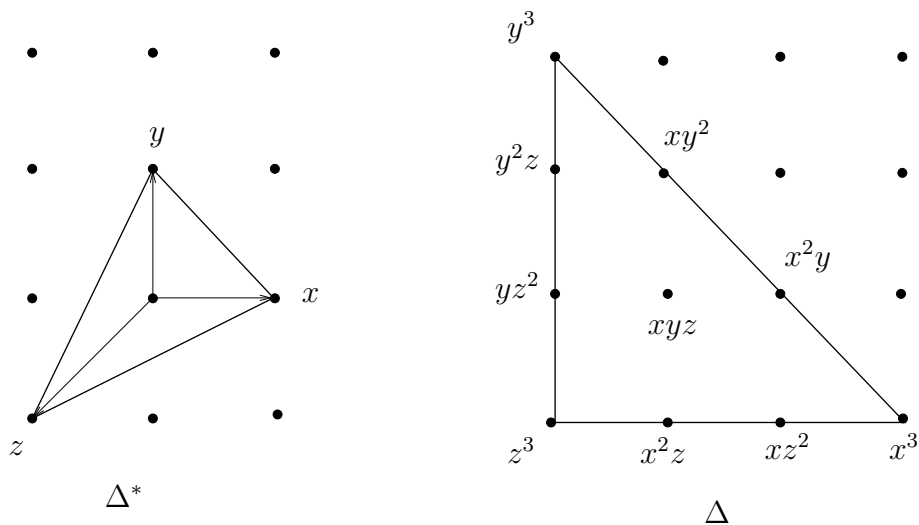}
\caption{The polytope representation of an elliptic curve as a hypersurface in $\IC \IP^2$.}
\label{f:hyper}
\end{center}
\end{figure}

\begin{example}
Let us give the example of a \CY\ hypersurface in $\IC \IP^2$, which is an elliptic curve. The complete triangulation of the polytope $\D^* \in N_\IR$ should give the fan of $\IC \IP^2$ presented in figure \ref{CP2}. As usual, we associate the homogeneous coordinates $x,y,z$ to the one-dimensional cones in the fan. This is shown in figure \ref{f:hyper}. It is now easy to compute the dual polytope $\D \in M_\IR$, to which we associate monomials as above. The family of hypersurfaces that we obtain is then given by the zero locus of cubic polynomials in $\IC \IP^2$, which is a well known way of describing an elliptic curve.
\end{example}

\subsubsection{\CY\ condition}

In fact, it was shown by Batyrev \cite{Batyrev:1994} that the hypersurface defined by the vanishing of a generic polynomial in the class determined by $\D$ is a smooth \CY\ manifold for $n \leq 4$, where $n$ is the dimension of the lattice $M$. For $n\leq3$ the underlying toric variety is smooth; in particular for $n=3$ the hypersurface describes a smooth K3 surface as explained above. For $n=4$ it may have point--like singularities, which are however missed by the generic hypersurface describing the \CY\ threefold.

Let us now explain why the hypersurface is a \CY\ manifold. Let a manifold $X$ be defined by the equation $P=0$ in a toric variety $\cm$. As we have seen in the first section, the polynomial $P$ defines a section of a line bundle (other sections are defined by different coefficients $a_j$). The divisor class of the line bundle can be read off from any monomial in $P$. Since the origin is always included in the polytopes, $P$ always includes the monomial $\prod_{i=1}^k w_i$, which corresponds to the divisor class $[\sum_{i=1}^k D_i]$. Thus, the polynomial $P$ determines a section of the anticanonical bundle of the toric variety $\cm$, which is crucial. 

We have seen in section 3 that in this case $P$ serves as a coordinate near $X$, and in fact the normal bundle $N_X$ of $X$ is simply $K^*_\cm|_X$, since $P$ is a section of the anticanonical bundle of $\cm$. Thus, the exact sequence $0 \to T^{1,0}X \to T^{1,0} \cm|_X \to N_X \to 0$ becomes
\beq
0 \to T^{1,0}X \to T^{1,0} \cm|_X \to K^*_\cm|_X \to 0.
\eeq
Now, given any holomorphic vector bundle $B$ over $X$ of rank $k$ and any holomorphic sub-bundle $A$, one can always form the respective determinant bundles $\det B$ and $\det A$ which satisfy the identity $\det B = \det A \otimes \det (B / A)$. Using the above exact sequence, we can then write
\beq
\det T^{1,0} \cm|_X = \det T^{1,0} X \otimes \det K^*_\cm|_X.
\eeq
Using the definition of the anticanonical bundle as the determinant line bundle of the holomorphic tangent bundle and the fact that $\det K^*_\cm = K^*_\cm$ since $K^*_\cm$ is a line bundle, we find
\beq
K^*_\cm|_X = K^*_X \otimes K^*_\cm|_X,
\eeq
or equivalently
\beq
K_X = (K^*_\cm \otimes K_\cm)|_X,
\eeq
that is the canonical bundle $K_X$ of $X$ is trivial, hence it is \CY.

\subsubsection{Fibration structure}

An interesting fact about \CY\ manifolds constructed as above is that their fibration structure (if any) can be read off directly from the reflexive polytopes. Let us explain how this works for an elliptically fibered K3 surface.

Take a three-dimensional pair of reflexive polytopes $\D$ and $\D^*$ describing a K3 manifold. Suppose that the intersection of $\D^*$ with the plane $\bar z = 0$ gives a reflexive polygon. We may reinterpret $P$ as a polynomial in the $w_i$ for which $\bar z_i=0$, with coefficients depending on the remaining $w_i$, i.e. we are dealing with an elliptic curve parameterized by the $w_i$ for which $\bar z_i\ne 0$. The map $\CM_{\Sigma} \to \IP^1$,
\beq
(w_1,\ldots, w_n) \;\to \; W=\prod_{i:\bar z_i\ne 0} w_i^{\bar z_i} 
\eeql{projmap}
is easily checked to be consistent with \refeq{equiv} and thus well defined. At any point of the $\IP^1$ that is neither $0$ nor $\infty$ all the $w_i$ with $\bar z_i\ne 0$ are non-vanishing, and \refeq{equiv} can be used to set all except one of them to 1. Hence, this gives the K3 surface the structure of an elliptic fibration.

It is easy to see that this kind of structure also holds for fibrations of higher-dimensional \CY\ varieties. Given a \CY\ manifold $X$ with a fibration structure such that the fiber is a lower-dimensional \CY\ manifold, the fiber is described by a subpolytope $\D^*_{\rm fiber}$ of the reflexive polytope $\D^*$ associated to $X$. See \cite{Skarke:1998yk} for more on this subject.

\subsubsection{Hodge numbers}

Another interesting property of the Batyrev's construction is that the Hodge numbers can be read off directly from the lattice data describing the \CY\ manifold. We will not give the explicit formulae here; the reader is referred to \cite{Batyrev:1994}. Owing to this fact, \CY\ hypersurfaces in toric varieties offer a fantastic playground to learn more about mirror symmetry, which exchanges the Hodge numbers $h^{2,1}$ and $h^{1,1}$ of a \CY\ threefold. In fact, mirror symmetry for hypersurfaces of toric varieties has been studied extensively and led to many interesting insights, as explained in \cite{Hori:2003}.

\providecommand{\href}[2]{#2}\begingroup\raggedright\endgroup


\begin{thebibliography}{10}



\bibitem{Aganagic:2003db}
M.~Aganagic, A.~Klemm, M.~Mari\~no, and C.~Vafa, ``The topological vertex,'' {\em
  Commun. Math. Phys.} {\bf 254} (2005) 425--478,
\href{http://www.arXiv.org/abs/hep-th/0305132}{{\tt hep-th/0305132}}.

\bibitem{Batyrev:1994}
V.~V. Batyrev, ``Dual Polyhedra and Mirror Symmetry for Calabi-Yau
  Hypersurfaces,'' {\em J. Alg. Geom.} {\bf 3} (1994) 493,
  \href{http://www.arXiv.org/abs/alg-geom/9310003}{{\tt alg-geom/9310003}}.

\bibitem{Bouchard:2003bu}
V.~Bouchard and H.~Skarke, ``Affine Kac-Moody algebras, CHL strings and the
  classification of tops,'' {\em Adv. Theor. Math. Phys.} {\bf 7} (2003)
  205--232,
\href{http://www.arXiv.org/abs/hep-th/0303218}{{\tt hep-th/0303218}}.

\bibitem{Bouchard:2005th}
V.~Bouchard, {\em Toric Geometry and String Theory}.
\newblock PhD thesis, University of Oxford, 2005,
\newblock \href{http://www.arXiv.org/abs/hep-th/0609123}{{\tt hep-th/0609123}}.

\bibitem{Bouchard:2005ag}
  V.~Bouchard and R.~Donagi,
  ``An SU(5) heterotic standard model,''
  Phys.\ Lett.\ B {\bf 633} (2006) 783,
  \href{http://www.arXiv.org/abs/hep-th/0512149}{{\tt hep-th/0512149}}.

\bibitem{Bouchard:2006}
  V.~Bouchard, M.~Cveti\v c and R.~Donagi,
  ``Tri-linear couplings in an heterotic minimal supersymmetric standard
  model,''
  Nucl.\ Phys.\ B {\bf 745} (2006) 62,
  \href{http://www.arXiv.org/abs/hep-th/0602096}{{\tt hep-th/0602096}}.
\bibitem{Calabi:1956}
E.~Calabi, ``The space of K\"ahler metrics,'' in {\em Proceedings of the
  International Congress of Mathematicians, Amsterdam, 1954, vol. 2},
  pp.~206--207.
\newblock North-Holland, Amsterdam, 1956.

\bibitem{Calabi:1957}
E.~Calabi, ``On K\"ahler manifolds with vanishing canonical class,'' in {\em
  Algebraic geometry and topology, a symposium in honour of S. Lefschetz},
  pp.~78--89.
\newblock Princeton University Press, Princeton, 1957.


\bibitem{Candelas:1987is}
P.~Candelas, ``Lectures on Complex Geometry,'' in {\em Trieste 1987,
  Proceedings, Superstrings '87}.
\newblock 1987.

\bibitem{Cox:1999}
D.~Cox and S.~Katz, {\em Mirror Symmetry and Algebraic Geometry}, vol.~68 of
  {\em Mathematical Surveys and Monographs}.
\newblock American Mathematical Society, 1999.
\newblock 469p.


\bibitem{Cox:1993fz}
D. Cox, ``The Homogeneous Coordinate Ring of a Toric Variety, Revised
  Version,''
\href{http://www.arXiv.org/abs/alg-geom/9210008}{{\tt alg-geom/9210008}}.

\bibitem{Fulton:1993}
W.~Fulton, {\em Introduction to Toric Varieties}.
\newblock Annals of Mathematics Studies. Princeton University Press, 1993.
\newblock 157p.

\bibitem{Greene:1996cy}
B.~R. Greene, ``String theory on Calabi-Yau manifolds,''
\href{http://www.arXiv.org/abs/hep-th/9702155}{{\tt hep-th/9702155}}.

\bibitem{Green:1987mn}
M.~B. Green, J.~H. Schwarz, and E.~Witten, {\em Superstring Theory. Vol. 2:
  Loop Amplitudes, Anomalies and Phenomenology}.
\newblock Cambridge Monographs On Mathematical Physics. Cambridge University
  Press, Cambridge, UK, 1987.
\newblock 596p.

\bibitem{Griffiths:1978}
P.~Griffiths and J.~Harris, {\em Principles of Algebraic Geometry}.
\newblock John Wiley \& Sons, 1978.
\newblock 813p.

\bibitem{Hori:2003}
K.~Hori, S.~Katz, A.~Klemm, R.~Pandharipande, R.~Thomas, C.~Vafa, R.~Vakil, and
  E.~Zaslow, {\em Mirror Symmetry}, vol.~1 of {\em Clay Mathematics
  Monographs}.
\newblock American Mathematical Society, 2003.
\newblock 929p.

\bibitem{Hubsch:1992}
T.~H\"ubsch, {\em Calabi-Yau Manifolds: A Bestiary for Physicists}.
\newblock World Scientific, 1992.
\newblock 374p.

\bibitem{Joyce:2000}
D.~D. Joyce, {\em Compact Manifolds with Special Holonomy}.
\newblock Oxford Mathematical Monographs. Oxford University Press, 2000.
\newblock 436p.


\bibitem{Marino:2004uf}
M.~Mari\~no, ``Chern-Simons theory and topological strings,''
\href{http://www.arXiv.org/abs/hep-th/0406005}{{\tt hep-th/0406005}}.

\bibitem{Nakahara:2002}
M.~Nakahara, {\em Geometry, Topology and Physics}.
\newblock The Institute of Physics, 2002.
\newblock 520p.


\bibitem{Neitzke:2004ni}
A.~Neitzke and C.~Vafa, ``Topological strings and their physical
  applications,''
\href{http://www.arXiv.org/abs/hep-th/0410178}{{\tt hep-th/0410178}}.

\bibitem{Rossi:2004}
M.~Rossi, ``Geometric Transitions,''
  \href{http://www.arXiv.org/abs/math.AG/0412514}{{\tt math.AG/0412514}}.


\bibitem{Skarke:1998yk}
H.~Skarke, ``String dualities and toric geometry: An introduction,''
\href{http://www.arXiv.org/abs/hep-th/9806059}{{\tt hep-th/9806059}}.

\bibitem{Yau:1977}
S.-T. Yau, ``On Calabi's conjecture and some new results in algebraic
  geometry,'' {\em Proceedings of the National Academy of Sciences of the
  U.S.A.} {\bf 74} (1977) 1798--1799.

\bibitem{Yau:1978}
S.-T. Yau, ``On the Ricci curvature of a compact K\"ahler manifold and the
  complex Monge-Amp\`ere equations. I.,'' {\em Communications on pure and
  applied mathematics} {\bf 31} (1978) 339--411.




\end{thebibliography}
\end{document}